\documentclass[galaxies,review,accept,oneauthor,pdftex]{Definitions/mdpi} 
\setitemize{parsep=6pt,itemsep=0pt,leftmargin=*,labelsep=5.5mm}
\setenumerate{parsep=6pt,itemsep=0pt,leftmargin=*,labelsep=5.5mm}
\setlist[description]{itemsep=0mm}   

\firstpage{1} 
\pubvolume{xx}
\issuenum{1}
\articlenumber{5}
\pubyear{2020}
\copyrightyear{2020}
\history{Received: 15 March 2020; Accepted: 20 April 2020; Published: date}
\updates{yes} 

\Title{Predictions and Outcomes for the Dynamics of Rotating Galaxies 
}

\newcommand{\LCDM}{$\Lambda$CDM}
\newcommand{\azero}{\ensuremath{\mathrm{a}_{0}}}
\newcommand{\ML}{\ensuremath{\Upsilon_*}}

\newcommand{\accunits}{\ensuremath{\mathrm{m}\,\mathrm{s}^{-2}}}
\newcommand{\MLsun}{\ensuremath{\mathrm{M}_{\odot}/\mathrm{L}_{\odot}}}
\newcommand{\Lsun}{\ensuremath{\mathrm{L}_{\odot}}}
\newcommand{\Msun}{\ensuremath{\mathrm{M}_{\odot}}}
\newcommand{\Vinf}{\ensuremath{V_{\infty}}}
\newcommand{\Vf}{\ensuremath{V_{f}}}

\Author{Stacy McGaugh\orcidA{}}

\AuthorNames{Stacy McGaugh}

\address[1]{%
Department of Astronomy, Case Western Reserve University, 10900 Euclid Ave., Cleveland, OH 44106, USA; ssm69@case.edu\\
}

\abstract{A review is given of {\textit{a priori}} %
 predictions made for the dynamics of rotating galaxies.
One~theory---MOND---has had many predictions corroborated by subsequent observations. 
While it is sometimes possible to offer post hoc 
explanations for these observations in terms of dark matter,
it is seldom possible to use dark matter to \textit{predict} the same phenomena.}


\begin{document}


\section{Introduction}

The dark matter problem remains unsolved after decades of intensive research. 
The observational evidence for mass discrepancies in extragalactic systems is overwhelming, 
but a laboratory detection of dark matter particles is still lacking.
While the need for ``dark matter'' is clear, its existence remains~hypothetical.

I place ``dark matter'' in quotes because the widespread use of this term presupposes the answer.
All that is really known is that the application of conventional dynamics to the visible mass distributions
of extragalactic systems fails to explain the observed motions. Though~this discrepancy is clear and the
 evidence is abundant~\cite{FG79,Trimble87,BigOtests}, much of this evidence is ambiguous as to whether the cause 
is unseen mass---literal dark matter---or a failure of the equations that lead to its inference~\cite{MdB98a,MdB98b,SMmond}.

At present, the~mainstream paradigm (the "normal component," as Feyerabend called it~\cite{Feyerabend}) 
is lambda cold dark matter (\LCDM). This paradigm works if and only if the cold dark matter (CDM) it presupposes
is a real, physical substance, and~not merely an abstraction that is convenient to cosmic calculations. 
This has motivated extensive laboratory searches for plausible dark matter candidates~\cite{Peebles1984,origWIMP,partphys_DM}.
These remarkable experiments~\cite{astrpart_searches,Xenon1T} have excluded essentially all of the parameter space in which 
the hypothesized dark matter particles were expected~\cite{Trotta2008} to reside.
This provides one motivation for considering ideas outside the normal component,
including new dark matter candidates and modified~dynamics.  

Another motivation arises if a theory makes successful predictions. 
Novel predictions that provide unique tests of hypotheses are the keystone of the scientific method.
The gold standard for scientific predictions are those made in advance of the observation~\cite{Merritt_convention}.
Here I highlight important \textit{a priori} predictions made by the modified Newtonian dynamics
(MOND) \cite{milgrom83a,milgrom83b,milgrom83c} that meet this gold standard.
I also discuss the contemporaneous expectation for dark~matter. 

I make no attempt to cover all aspects of MOND and the mass discrepancy problem here, as the subject is now vast.
More extensive reviews of MOND are provided elsewhere
~\cite{milgromrev1999,milgromrev2001,bekenstein,milgromrev2008,LivRev,milgromrev2014,CJP},
as are reviews of the relevant data~\cite{S1990,S1996,MdB98b,SMmond,M06,IAU_review}.
Here I focus on tests of \textit{a priori} predictions utilizing the most accurate data that relate to the subject~\cite{IAU_review,SPARC}.

\section{Predictions and~Tests}
\label{sec:pred}

General relativity has been tested with extraordinary precision in the solar system~\cite{Will2014}, and high acceleration systems, such as binary neutron stars~\cite{binarypulsar} and merging black holes~\cite{LIGO}. In~contrast, it~manifestly fails in systems that
exhibit mass discrepancies; hence the need for dark matter.
Attempts to modify dynamics often start by noting that problem systems---galaxies,
clusters of galaxies, and~the universe as a whole---are much bigger than the systems where established theory works so well.
Consequently, it is tempting to imagine that the force law changes on some length scale 
$R_{gal} \approx 1\;\mathrm{kpc}$ so that its effects are imperceptible in the solar system but pronounced in galaxies.
This approach immediately runs afoul of the observation that some large galaxies appear to require little dark matter, while some
small galaxies evince large discrepancies. Modifications based on a length scale can be generically excluded~\cite{MdB98a}.

Size is not the only scale that sets problematic systems apart from solar system tests of gravity. The~typical accelerations of
stars in galaxies are of order 1 \AA\ s$^{-2}$ or less; this is eleven orders of magnitude less than we experience on the surface
of the Earth, and~many orders of magnitude removed from sensitive solar system probes.
MOND~\cite{milgrom83a} hypothesizes a change to the effective force law at low accelerations,
$a < \azero$. The~acceleration scale \azero\ is empirically determined to be $\azero = 1.2 \times 10^{-10}\;\accunits$~\cite{BBS}.
The value of \azero\ has been remarkably stable, having not changed meaningfully in decades~\cite{RAR}. 

As noted in the original publication~\cite{milgrom83a}, MOND is not a complete theory that replaces general relativity.
Indeed, MOND may be either a modification of gravity (Newton's universal gravitation) or a modification
of the law of inertia ($F = ma$: the inertial mass may differ from the gravitational charge at low accelerations) \cite{milgrom83a,milgromrev2014}.
Perhaps it is only an effective theory that arises for reasons we have yet to imagine.
Irrespective of why it happens, strong predictions follow once a force is~hypothesized.

MOND contains Newton in the limit of high acceleration: for $a \gg \azero$, the~effective acceleration
$a = g_N$, where $g_N$ is the usual Newtonian gravitational
force per unit mass. Everything is "normal" until we reach the regime of low acceleration ($a \approx \azero$); an immediate corollary is that
the need for dark matter should never appear at high accelerations. Unique predictions of
MOND emerge in the deep MOND regime: for $a \ll \azero$, the~effective acceleration becomes $a = \sqrt{g_N \azero}$. 
Intriguingly, dynamics become scale invariant in this deep MOND regime~\cite{ScaleInvar}.
The Newtonian and deep MOND regimes are connected by a theoretically arbitrary but empirically well-constrained 
interpolation function whose details are only relevant within a factor of a few of \azero. 
The essence of the idea is captured by the asymptotic limits at high and low accelerations, which is where the important predictions~arise.

In the following, I review observational tests of the specific predictions 
elaborated in section VIII of~\cite{milgrom83b} that have been subsequently~tested.

\subsection{Tully--Fisher and the Mass--Asymptotic Speed~Relation}

\begin{quote}
"The $\Vinf^4 = \azero G M$ relation should hold~exactly." \\
---M.\ Milgrom~\cite{milgrom83b}
\end{quote}

One consequence of MOND is a relation between the mass of a galaxy and its rotation speed.
One immediately recognizes this mass--asymptotic speed relation (MASR) \cite{milgromrev2014}
as the basis of the empirical Tully--Fisher relation~\cite{TForig},
\begin{equation}
L \sim W^x,
\label{eq:TF}
\end{equation}
provided that luminosity is a proxy for mass ($L \sim M$) and line-width is a proxy for rotation speed ($W \sim \Vinf$).
The Tully--Fisher relation provides several tests of~MOND.

{Testing the MASR requires careful measurement of both the mass and the asymptotic speed.}

Many rotation curves are observed to be flat, but~it sometimes happens that the observational sensitivity tapers off before 
the asymptotic rotation speed is obtained. 
One must therefore take care to test the theory and not just the limits of data quality~\cite{IAU_review}:
the result will differ from the prediction if either of the proxies for mass or \Vinf\ are imperfect~\cite{Lelli19}. 

The flat rotation velocity \Vf\ measured from resolved rotation curves provides a better proxy for \Vinf\ than line-widths.
It is still only a proxy, as~rotation curves in MOND may approach a constant rotation speed quickly, but~may also
decline slowly or rise gradually depending on the details of the mass distribution~\cite{milgrom83b}.
This morphology is clearly seen in the data~\cite{OneLaw,IAU_review}. Nevertheless, it is often 
possible\footnote{It is also possible to mistakenly conclude that MOND is incorrect~\cite{superspirals} by utilizing an 
inadequate proxy for \Vf\ \cite{Noordetal,noordTF}.} to measure \Vf\ to within 5\% \cite{SPARCTF,Lelli19}. 

Another important effect is geometric: flattened mass distributions like spiral galaxies
rotate faster than the spherically equivalent distribution~\cite{BT}.
The MASR predicts a Tully--Fisher-like relation of the form
\begin{equation}
M = \frac{\zeta \Vf^4}{\azero G}
\label{eq:BTFR}
\end{equation}
where $\zeta$ is a factor of order unity that accounts for the
flattened geometries of disk galaxies. This can be computed analytically for a razor thin exponential disk (see Equation~(16) of~\cite{MdB98b}),
with the result that $\zeta = 0.76$ at four disk scale lengths. For~disks of realistic finite thickness, $\zeta \approx 0.8$ \cite{M05}. 
As a practical matter, this quantity likely has some intrinsic scatter~\cite{CDR}, and~may vary systematically with mass or morphological type.
Systematic variation would affect the slope of the~BTFR.

Empirically, the~data evince a baryonic Tully--Fisher relation (BTFR) \cite{btforig}
\begin{equation}
M =A \Vf^x.
\end{equation}

Here $M$ includes all relevant forms of baryonic mass: {stars, their remnants, all phases of interstellar gas, and~dust.
In practice, the~dominant forms of baryonic mass in late type galaxies are stars (including the corresponding remnants) and atomic gas.
We estimate the baryonic mass using stellar population models \cite[][]{SM14,MS14} to estimate mass-to-light ratios \ML\ and 
near-IR luminosities~\cite{SM14Spitzer,SPARC} in the $3.6\mu$ band of the Spitzer Space Telescope so that $M_* = \ML L_{[3.6]}$. 
The gas mass is estimated from the atomic gas mass corrected for the hydrogen 
fraction (see \cite{MLSRNAAS2020}) so that the total baryonic mass is} $M = M_* + M_g$.

{The BTFR} is equivalent to the MASR \textit{if} \Vf\ is an adequate proxy for \Vinf, the~slope $x = 4$, \textit{and} the normalization $A$ is
consistent with $A = \zeta/(\azero G)$ for realistic galaxy masses.
Another important implication of the MASR that follows from MOND is that it is only
the baryonic mass of a galaxy that sets its {asymptotic} rotation speed, not $M/r$ as in Newtonian dynamics. 
We discuss these distinct aspects of the MOND prediction for the BTFR in~turn.

\begin{Property} 
MASR Normalization
\label{pr:norm}
\end{Property}

The normalization of the MASR predicted by MOND is determined by 
fundamental constants: $(\azero G)^{-1} = 63\;\Msun\,\mathrm{km}^{-4}\,\mathrm{s}^4$.

\paragraph{\textit{Do the data corroborate the prediction of MOND?}} Yes.
For finite thickness disk galaxies with $\zeta = 0.8$, the~prediction of the MASR
corresponds to a BTFR with $A = 50\;\Msun\,\mathrm{km}^{-4}\,\mathrm{s}^4$.
This is consistent with the available data for rotationally supported galaxies~\cite{verhTF,M05,noordTF,stark,trach,M11,M12,SPARCTF,Lelli19}.

\paragraph{\textit{Was the prediction made} a priori?} No. 
This is a good test of MOND, but~one has to appeal to data to set the value of \azero\ in the first place. Consequently, this test
of MOND is successful, but~does not meet the gold standard of an \textit{a priori}~prediction.

\paragraph{\textit{What does dark matter predict?}} The expectation for the normalization of the Tully--Fisher relation in \LCDM\ was discussed
at length in~\cite{M12}. Nominally, one expects a higher normalization than observed~\cite{SN,NS,MayerMoore}, in~the sense that the are more
baryons available in dark matter halos to form stars than apparently do so~\cite{M10}. One of the primary reasons for invoking highly
efficient feedback in more recent numerical simulations is to prevent the otherwise inevitable cooling and subsequent formation into stars
of these excess~baryons.

\begin{Property} 
MASR Slope
\label{pr:slope}
\end{Property}

\begin{figure}[h]
\centering
\includegraphics[width=6 in]{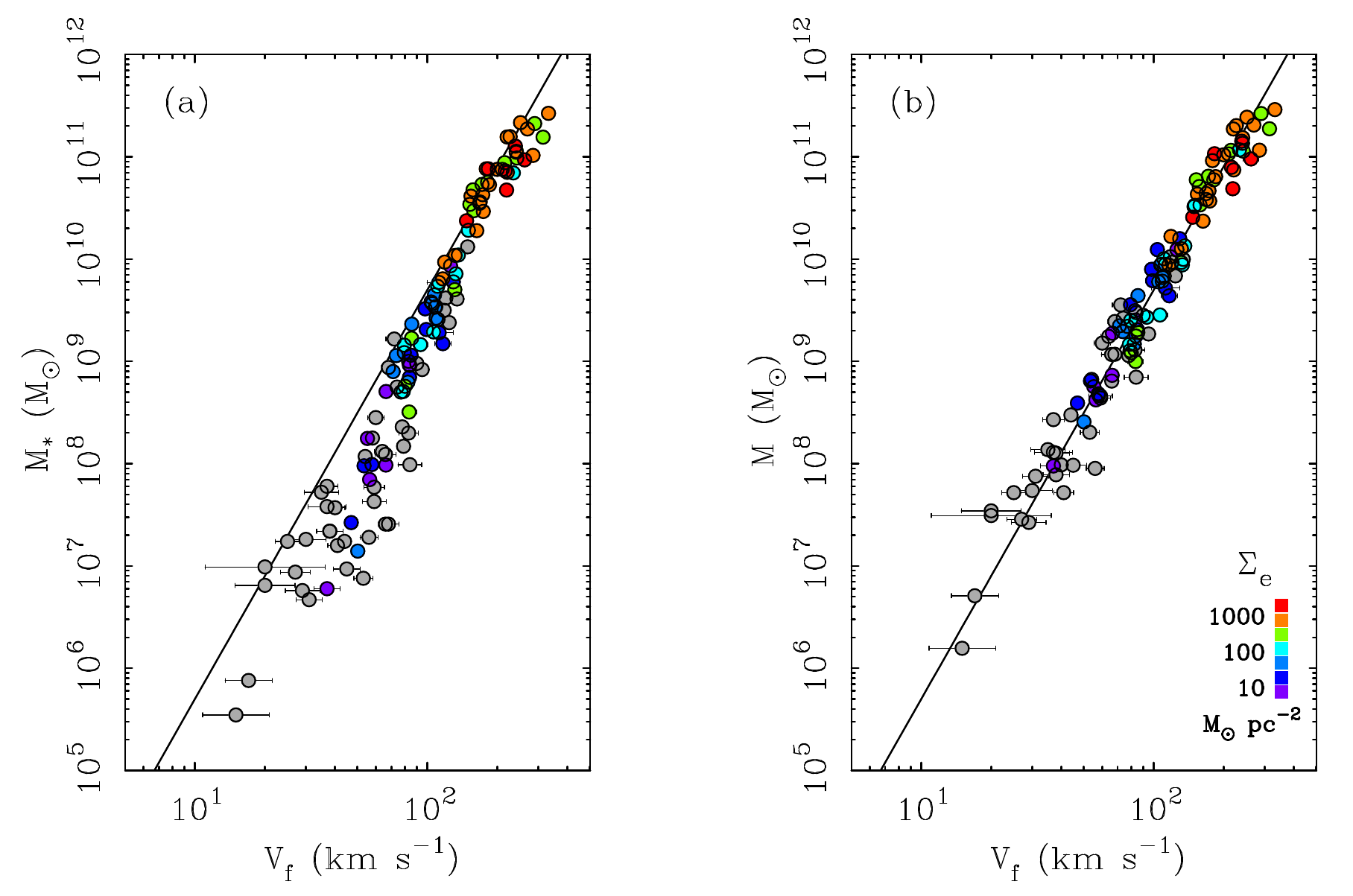}
\caption{Tully--Fisher relations: the flat rotation speed \Vf\ as a function of (\textbf{a}) stellar mass, and~(\textbf{b}) baryonic
mass ($M = M_*+ M_g$). Gas masses follow from observed 21cm fluxes~\cite{ISMbook}. Stellar masses are estimated from observed
luminosities and a population synthesis prescription for the stellar mass-to-light ratio: $M_* = \ML L$ \cite{MS14}.
Here, and~throughout the paper unless otherwise noted, we adopt $\ML = 0.5\;\MLsun$ for star forming disks and 
$0.7\;\MLsun$ for bulge components in the $3.6\,\mu$ band (often called [3.6]) of the Spitzer Space Telescope~\cite{SM14,meidt,SML19}.
Rotation speeds are obtained from resolved rotation curves that are extended enough to measure \Vf\ \cite{SPARCTF,Lelli19,M12}. 
Points are color coded by the effective stellar surface density when known 
from Spitzer data~\cite{SM14Spitzer,SPARC}, ranging from low (blue) to high (red) surface brightness (scale inset).
The gray points are gas dominated low surface brightness (LSB) galaxies that meet the quality criteria discussed by~\cite{M12} but do not have Spitzer data.
The line is the prediction of MOND~\cite{milgrom83b} for the value of \azero\ found by~\cite{BBS} before these data existed.
\label{fig:BTFR}}
\end{figure}

\paragraph{\textit{Do the data corroborate the prediction of MOND?}} Yes. The~data are consistent with the predicted slope $x=4$.
Figure~\ref{fig:BTFR} shows the data reviewed by~\cite{IAU_review} along with the MOND prediction.
The line representing MOND has not been fit. It has a slope $x=4$ and a normalization $A = 50\;\Msun\,\mathrm{km}^{-4}\,\mathrm{s}^4$,
as discussed above~\cite{M05}. Clearly, these data are consistent with the prediction of~MOND. 

An important consideration in constraining the slope of the BTFR is the dynamic range in the data.
Observational selection effects severely bias galaxy samples in favor of high luminosity, high~surface brightness (HSB) galaxies
and against low luminosity and low surface brightness (LSB) galaxies~\cite{MBS95,IAU_review}.
Consequently, most realizations of the Tully--Fisher relation are dominated by high mass galaxies, and~sample only a narrow range
in mass: the bulk of most data sets are confined to spiral galaxies with $10^{10} < M_* < 2 \times 10^{11}\;\Msun$, typically with only
a few galaxies down to $\sim$$10^9\;\Msun$. 
This results in a systematic underestimate of the fitted slope~\cite{AkritasBershady}.
As the dynamic range over which data are available has expanded ($M \rightarrow 10^7\;\Msun$) \cite{stark,trach,begum,MS15}, 
the slope steepens and $x \rightarrow 4$ \cite{MS15}.

\paragraph{\textit{Was the prediction made} a priori?} Yes and no. 
The Tully--Fisher relation was known prior to the development of MOND~\cite{TForig,AaronsonTF},
such~that it was not an \textit{a priori} prediction. The~slope was highly uncertain when MOND was developed: 
Milgrom quotes the range $2.5 < x < 5$ \cite{milgrom83b}. A~slope of 4 is within that range, but~there was no guarantee
that the data would settle on that~value.

An important consequence of the slope 4 MASR of MOND is the location of low mass galaxies in the BTFR plane~\cite{M11,MWolf}.
Rotation speeds of the low mass galaxies represented by grey points in~Figure~\ref{fig:BTFR} were not known at the time MOND was hypothesized.
Yet it was possible to use MOND to successfully predict the rotation speeds of these objects. This prediction follows directly
from the statement of the MASR. Its specific application in this context was discussed by~\cite{M05} and subsequently applied by~\cite{M11}. 
This does constitute a successful \textit{a priori} prediction.

\paragraph{\textit{What does dark matter predict?}} CDM\ does not make a specific prediction for the BTFR. It does predict a
mass--rotation speed relation for dark matter halos: $M_{200} \sim V_{200}^3$ \cite{MMW98,SN}. To~connect this to the BTFR, 
we must introduce proportionality factors $m_d$ and $f_v$ that relate the observed baryonic mass to the total mass 
$M = m_d M_{200}$ and the observed rotation speed to that predicted at the virial radius $\Vf = f_v V_{200}$ \cite{M12}.
These~necessary proportionality factors are not cleanly predicted by galaxy formation models. The~obvious assumption
is that they be constants~\cite{MMW98}; this predicts a slope $x=3$ that is inconsistent with the observations.
Various effects during galaxy formation may steepen the slope~\cite{BullockTF,gnedinTF}, but~typically only to \mbox{$x \approx 3.4$}.
It often happens that \LCDM\ models~\cite{TGKPR} induce curvature (variable $x$) in the BTFR that is not observed (Figure \ref{fig:BTFR}).
Indeed, it has become difficult to avoid such curvature given the shape of the abundance matching relation~\cite{BBK}.
One can tune models to impose a slope of 4~\cite{vdB00}, but~then one has a fine-tuned model which is not satisfactory.
This approach has a propensity to violate other constraints. For~example, one can vary $m_d$
in the models of~\cite{MMW98} to obtain the desired BTFR slope. The~required variation ($m_d \propto \Vf$) then ruins the otherwise good
agreement with the disk size--mass relation that is obtained with constant $m_d$ \cite{MMW98}. One cannot have it both ways;
one property can be fit, but not both. More elaborate models can be constructed with more parameters, but~these violate Occam's
rule of parsimony, and~inevitably lack predictive power: they chase the data rather than predict it~\cite{Sanders_Aachen}. 

\begin{Property} 
Baryonic Mass and Flat Velocity
\label{pr:masr}
\end{Property}

A fundamental prediction of MOND is that the physical basis of the Tully--Fisher relation is a relation between flat rotation speed
and baryonic mass. All the normal mass matters. It does not matter whether the mass is in the form of a star or~gas. 

\paragraph{\textit{Do the data corroborate the prediction of MOND?}} Yes.
The BTFR (Figure \ref{fig:BTFR}) is a direct consequence of the MASR in~MOND.

\paragraph{\textit{Was the prediction made} a priori?} Yes. 
The absolute nature of the MASR was emphasized in the original papers~\cite{milgrom83b}.
The importance of gas and stars in this context was not widely appreciated until much~later. 

\paragraph{\textit{What does dark matter predict?}} I am not aware of any dark matter models that addressed 
this aspect of the Tully--Fisher relation prior to the empirical identification of the BTFR~\cite{btforig}. This is unsurprising,
since the BTFR is something of a non-sequitur in CDM: dark matter plays no direct role in its construction.
It is often assumed that \Vf\ is set by the dark matter halo, but~this presupposition is inadequate,
as the baryons make a non-negligible contribution to \Vf\ in HSB galaxies (see below). 

\begin{Property} 
Surface Brightness Independence
\label{pr:sbi}
\end{Property}

\begin{quote}
"Disk galaxies with low surface brightness provide particularly strong~tests." \\
---M.\ Milgrom~\cite{milgrom83b}
\end{quote}

An important consequence of the absolute nature of the MASR is that there should be no residuals from the BTFR.
The only variables that appear are the total baryonic mass and the flat rotation speed.
Neither size nor surface brightness appear in the equation, so there should be no dependence on these~quantities.

\paragraph{\textit{Do the data corroborate the prediction of MOND?}} Yes.
The absence of surface brightness residuals was recognized in the mid-90s by several independent 
groups~\cite{zwaanTF,sprayTF,hoffmanTF} and has been confirmed many times since.
Galaxies of different surface brightness all fall on the same BTFR (Figure \ref{fig:BTFR}).

\paragraph{\textit{Was the prediction made} a priori?} Yes. This was explicitly predicted~\cite{milgrom83b}:
"We predict, for~example, that the proportionality factor in the $M \propto \Vinf^4$ relation for (LSB) galaxies
is the same as for the high surface density galaxies."

\paragraph{\textit{What does dark matter predict?}} Conventionally, 
it had been expected that LSB galaxies should shift off of the Tully--Fisher relation defined by 
HSB galaxies~\cite{AaronsonTF,milgrom83b}, since the rotation speed depends on
size as well as mass: $V^2 \sim M/r$. 
By squaring this, we obtain $V^4 \sim L \Sigma$.
It was argued~\cite{AaronsonTF} that a Tully--Fisher relation of the form $V^4 \sim L \Sigma$ follows if the surface brightness $\Sigma$ is
the same for all galaxies, as~was then believed~\cite{F70}. This~fails when confronted with data for LSB galaxies~\cite{zwaanTF},
which have different $\Sigma$ by~definition.

The absence of the anticipated residuals poses a fine-tuning problem for conventional dynamics~\mbox{\cite{MdB98a,CR}}.
The observed flat rotation speed is the sum of a declining luminous contribution and increasing dark contribution:
$\Vf^2 = V_*^2(R)+V_g^2(R)+V_{DM}^2(R)$. Galaxies span a wide range of surface brightness at a given mass, 
but are indistinguishable to Tully--Fisher~\cite{dBM96,TVbimodal,MdB98a,MdB98b,IAU_review}. As~surface brightness declines
at fixed mass, $V_*(R)$ declines with it, so $V_{DM}(R)$ must increase to precisely compensate and keep \Vf\ unchanged.
The only way to avoid this fine-tuning is if all galaxies are dark matter dominated~\cite{CR} so that 
$V_*(R) \ll V_{DM}(R)$ at all relevant radii~\cite{CR,myPRL}. 
This limit requires implausibly low stellar masses~\cite{StarkmanMaxDisk}, and~is directly contradicted by the
observed dependence of rotation curve shape on 
surface brightness~\cite{IAU_review,dBM96,TVbimodal,Swaters09,Swaters12,LelliVRgrad,CDR,diversity}.

\subsection{Predictions for Rotation~Curves}

\begin{quote}
"Rotation curves calculated on the basis of the observed mass distribution and the modified dynamics 
should agree with the observed velocity~curves." \\
---M.\ Milgrom~\cite{milgrom83b}
\end{quote}

This simple statement has a variety of testable~consequences.

\begin{Property} 
Flat Rotation Curves
\label{pr:flat}
\end{Property}

The striking flatness of the rotation curves of spiral galaxies~\cite{vera,bosma} was an animating motivation for both dark matter and MOND. 
That they are observed to be so is thus not an \textit{a priori} prediction. It is nevertheless a test: one should not observe galaxies that
show a Keplerian decline. So far, rotation curves remain flat indefinitely far out~\cite{THINGS,Brouwer17}.

\paragraph{\textit{Do the data corroborate the prediction of MOND?}} Yes.

\paragraph{\textit{Was the prediction made} a priori?} No: flat rotation curves were the motivation for MOND, not a prediction thereof.
The theory takes flat rotation curves to be axiomatic, an~expectation that could be falsified but has not been~\cite{THINGS}.
 
\paragraph{\textit{What does dark matter predict?}} Flat rotation curves were a primary motivation for dark matter, not a prediction thereof. 
It is generally possible to fit a variety of dark matter halos to the data, once given~\cite{K87,SPARChalos}. It is another matter to
predict rotation curves \textit{a priori}. It is easy to build plausible-seeming models with rotation curves that are not as flat as
those observed---indeed, it is hard to avoid~\cite{MdB98a}. Models with realistic rotation curve shapes are restricted to an 
unnaturally narrow range of the available parameter space~\cite{adiabat,Disney2008,DMH}. 


\begin{Property}
The Acceleration Discrepancy
\label{pr:accD}
\end{Property}

A straightforward property to compute for a galaxy is the enclosed dynamical mass-to-light ratio.
Assuming a spherical mass distribution, the~dynamical mass enclosed within radius $r$ is simply $M_{dyn}(< r) = r V^2/G$.
This may be compared to the luminosity or baryonic mass enclosed by the same radius, giving some idea of the amount of dark matter required.
This can be quantified by the mass discrepancy~\cite{S1990,M1999ASPC,M04}, which is the ratio of the observed centripetal acceleration
to that predicted by the observed baryons: $D = a/g_N \approx M_{dyn}/M_b$. The~equation with the ratio of dynamical to baryonic mass
is not exact because spiral galaxies are not spherical. Consequently, a~more accurate name would be the 
\textit{acceleration discrepancy} \cite{bek_accdisc}.

In MOND, the~amplitude of the discrepancy depends on the distribution of luminous mass.
If~we interpret this in terms of conventional dynamics, we should find that the enclosed dynamical mass-to-light ratio
varies predictably~\cite{milgrom83b}. Specifically, there should be no discrepancy when accelerations are above the
critical value \azero. That is, the~dynamical mass-to-light ratio should be comparable to that expected for the stellar
population mass-to-light ratio (typically of order unity in solar units, depending on pass-band~\cite{MS14}).
A transition should occur at $r_M \approx V^2/\azero$, after~which the discrepancy should increase
with increasing radius as the acceleration declines (for a flat rotation curve, $a \sim r^{-1}$).
The transition radius should vary systematically from galaxy to galaxy: it is \azero\ that is constant.
Consequently, the~discrepancy should be larger and set in at smaller radii in galaxies of lower surface brightness,
which are predicted to have lower~accelerations.

\begin{figure}[h]
\centering
\includegraphics[width=6 in]{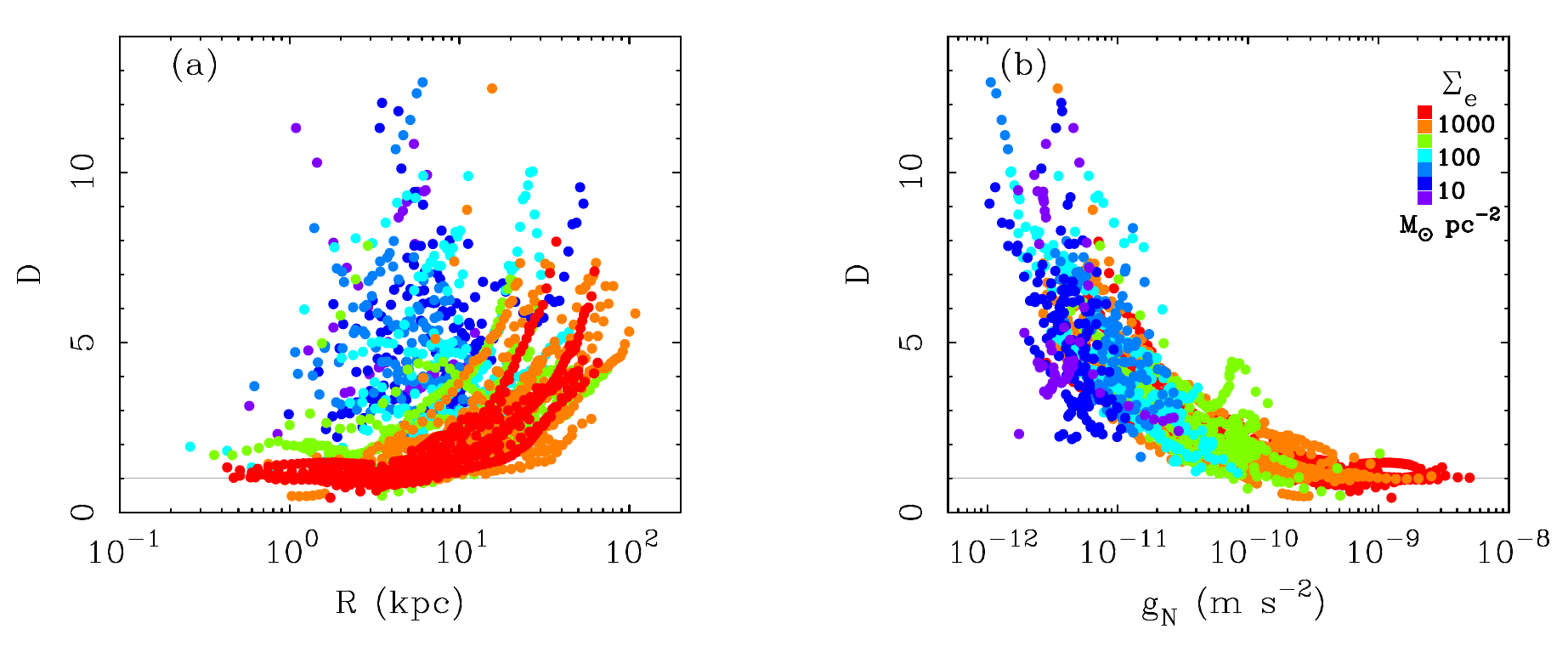}
\caption{The amplitude of the acceleration discrepancy $D = a/g_N$ of galaxies in the SPARC 
database~\cite{SPARC} as a function of (\textbf{a}) radius and (\textbf{b}) the expected acceleration.
Each point is an accurate ($\sigma_V/V < 5\%$), resolved measurement, with~multiple measurements per galaxy. 
The centripetal acceleration $a = V^2/R$ is measured from rotation curves while that predicted for the observed baryons $g_N$ 
is computed by solving the Poisson equation for the observed distribution of stars and gas. 
Points are color coded by the effective stellar surface densities of their galaxies (legend), as~in Figure~\ref{fig:BTFR}. 
The need for dark matter ($D > 1$) appears gradually at large radii in high~surface brightness (HSB) galaxies but is greater and sets in at smaller radii in 
galaxies of progressively lower surface brightness (\textbf{a}). This behavior was anticipated by MOND~\cite{milgrom83b},
along with the correlation of the amplitude of the discrepancy with acceleration (\textbf{b}).
\label{MLSB}}
\end{figure}

Indeed, the~point of MOND is that the mass discrepancy is an acceleration-dependent phenomenon. 
Hence the amplitude of the discrepancy $D$ should correlate with acceleration. This is apparent in Figure~\ref{MLSB}b,
a purely empirical correlation that has been known for a long time~\cite{S1990,M1999ASPC,M04} and which has become
especially clear with the availability of near-IR data from Spitzer~\cite{RAR}.  

\paragraph{\textit{Do the data corroborate the prediction of MOND?}} Yes. 
All aspects of the prediction are apparent in Figure~\ref{MLSB} (see also Figure~3 of~\cite{MdB98b}).
In the highest surface brightness spiral galaxies, the~accelerations are close to \azero\ at small radii, and~there is little 
indication of a dynamical discrepancy. The~discrepancy appears gradually as one goes out in radius, as~seen in a steadily
increasing dynamical mass-to-baryonic mass ratio. As~we consider galaxies of
progressively lower surface brightness, the~discrepancy appears sooner, at~smaller radii, and~is also larger in amplitude.
An important empirical point is that it is surface brightness, not luminosity~\cite{MdB98a,thirdlaw}, that determins $g_N$ and drives the 
correlation with $D$ \cite{S1990,M1999ASPC,M04,RAR}.

\paragraph{\textit{Was the prediction made} a priori?} Yes. 
This prediction was first explicitly tested~\cite{MdB98b} some 15 years after it was published~\cite{milgrom83b}.
It has become increasingly clear as the data have improved~\cite{RAR}.

\paragraph{\textit{What does dark matter predict?}} {I am not aware of an} explicit prediction {having been} 
made for this observation {in the context of dark matter}.
It is now widely known that LSB galaxies tend to be dark matter dominated, but~that is a recognition
driven entirely by the data~\cite{dBM96,dBM97}. There was no reason to expect this to be the case \textit{a priori}.
MOND was the only theory to correctly predict this behavior in advance of its~observation.

\begin{Property} 
Rotation Curves Shapes
\label{pr:shape}
\end{Property}

\begin{quote}
"The rotation curve of a galaxy can remain flat down to very small radii, as~observed, only if the galaxy's average surface density $\Sigma$ 
falls in some narrow range of values which agrees with the Fish and Freeman laws. For~smaller $\Sigma$, the~velocity rises more slowly to
the asymptotic~value." \\
---M.\ Milgrom~\cite{milgrom83b}
\end{quote}

\paragraph{\textit{Do the data corroborate the prediction of MOND?}} Yes. 
In MOND, the~shapes of rotation curves follow from their baryonic mass distributions. 
Bright, high surface brightness spirals are predicted to have steeply rising rotation curves that flatten quickly, or~even decline
before flattening. Low surface brightness galaxies should have slowly rising rotation curves that only gradually 
approach the flat velocity. Precisely this morphology is observed~\cite{MdB98b,LivRev,OneLaw,IAU_review}.

\paragraph{\textit{Was the prediction made} a priori?} Yes.
Flat rotation curves were known at the time that MOND was developed. However, the~overall \textit{shapes} of rotation curves
were only beginning to be explored. Milgrom's quote above nicely summarizes the state of knowledge at that time. 
Rotation curves that remain flat to small radii can only occur in MOND for HSB galaxies---hence
his explicit comment about a galaxy's average surface density. The~rotation curves of LSB galaxies were essentially unknown.
Indeed, at~the time, it was widely believed that rotationally supported galaxies all had essentially the same surface brightness
(Freeman's Law~\cite{F70}), and~LSB galaxies did not exist. Hence it is remarkable that an explicit prediction was made 
for LSB galaxies, let alone that this prediction was realized by subsequent observations~\cite{MdB98b,dBM98}.
 
\paragraph{\textit{What does dark matter predict?}} There are many schools of thought as to what should happen with LSB galaxies,
once they were recognized to exist. These fall into two broad categories~\cite{MdB98a}. In~one, it was imagined that LSB galaxies
were stretched out versions of HSB galaxies, residing in late-forming dark matter halos that were themselves of lower average
density. This hypothesis is rejected by the absence of surface brightness residuals in the BTFR (Figure \ref{fig:BTFR}), as~LSB
galaxies should have lower overall rotation speeds simply because $V^2 \sim M/r$, and by~construction, they have larger radii
at a given~mass.

\begin{figure}[h]
\centering
\includegraphics[width=6 in]{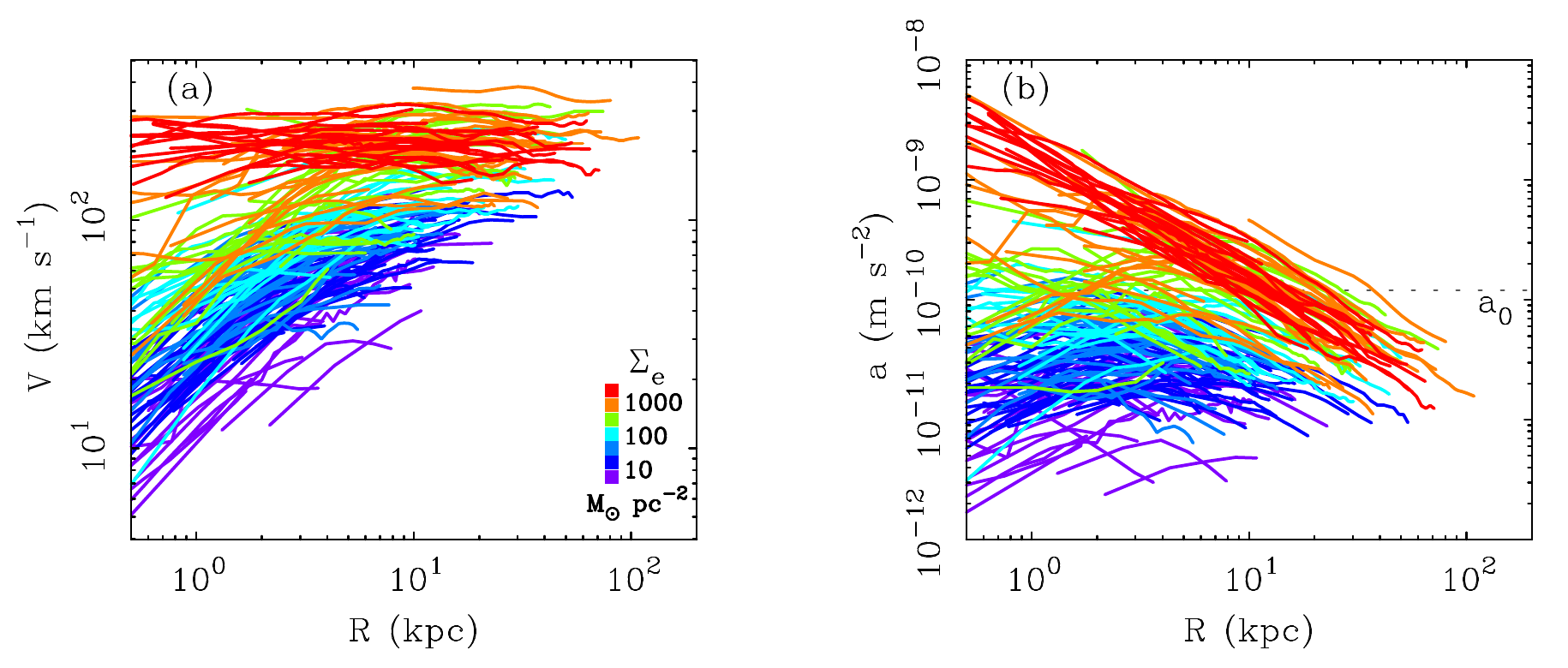}
\caption{The (\textbf{a}) rotation curves of galaxies in the SPARC database~\cite{SPARC}
and (\textbf{b}) the corresponding centripetal acceleration curves $a = V^2/R$.
Each line is one galaxy; the MOND acceleration scale \azero\ is noted in (\textbf{b}).
Galaxies are color coded by the effective stellar surface densities, as in Figures~\ref{fig:BTFR} and \ref{MLSB}.
High surface brightness spirals have rotation curves that rise sharply and flatten quickly. These are Freeman disks
like those known at the time that MOND was developed. In~contrast, LSB galaxies were essentially unknown at that time.
They were subsequently observed~\cite{MdB98b} to have slowly rising rotation curves that only gradually turn over and
approach the flat velocity, as~predicted by MOND~\cite{milgrom83b}. The~direct connection between stellar surface density
and dynamical acceleration predicted by MOND is illustrated by the rainbow variation in (\textbf{b}): low surface brightness
galaxies have low accelerations (often well below \azero), while high surface brightness galaxies have high accelerations.
\label{VRSB}}
\end{figure}

A more persistent school of thought is that galaxies of the same stellar mass reside in halos of the same total mass. 
Size follows from the initial angular momentum of the parent dark matter halo~\cite{FE_AM}, and~
the lack of BTFR residuals with surface brightness can be explained if and only if the stellar disk is sufficiently sub-maximal that
it does not impact \Vf. However, if~this is the case, then the rotation curve is dominated by dark matter halos, which are expected
to be very self-similar at a given mass~\cite{NFW}. This predicts that galaxies of the same mass have not only the same \Vf, but~that the entire shape of the rotation curve $V(R)$ should be very nearly the same. This expectation is not realized; there is a
greater diversity of observed rotation curve shapes~\cite{IAU_review,dBM96,TVbimodal,Swaters09,Swaters12,LelliVRgrad,CDR} 
than is predicted by such models~\cite{diversity}. This diversity is obvious in Figure~\ref{VRSB}: LSB galaxies have slowly rising
rotation curves, and HSB galaxies have rapidly rising rotation curves, just as predicted by MOND: the distribution of baryons matters as does their total mass. The~hypothesis that $V(R)$ should be essentially the same for galaxies of the same mass---still present
in some modern galaxy formation simulations~\cite{diversity}---is rejected by the~data.

\begin{Property} 
Surface Density Follows from Surface Brightness
\label{pr:sdens}
\end{Property}

In MOND, the~dynamical surface density should follow from the surface density of baryonic mass. 
This surface density of stars is well traced by the surface brightness in the near-infrared 
(e.g.,~the~$K$-band at $2.2\,\mu$ or the $3.6\,\mu$ band of the Spitzer Space Telescope).
The mass surface density is traced by the dynamics: $a \sim 2 \pi G \Sigma$. 
Consequently, we expect a correlation between surface brightness and measured~acceleration.

\begin{figure}[h]
\centering
\includegraphics[width=6 in]{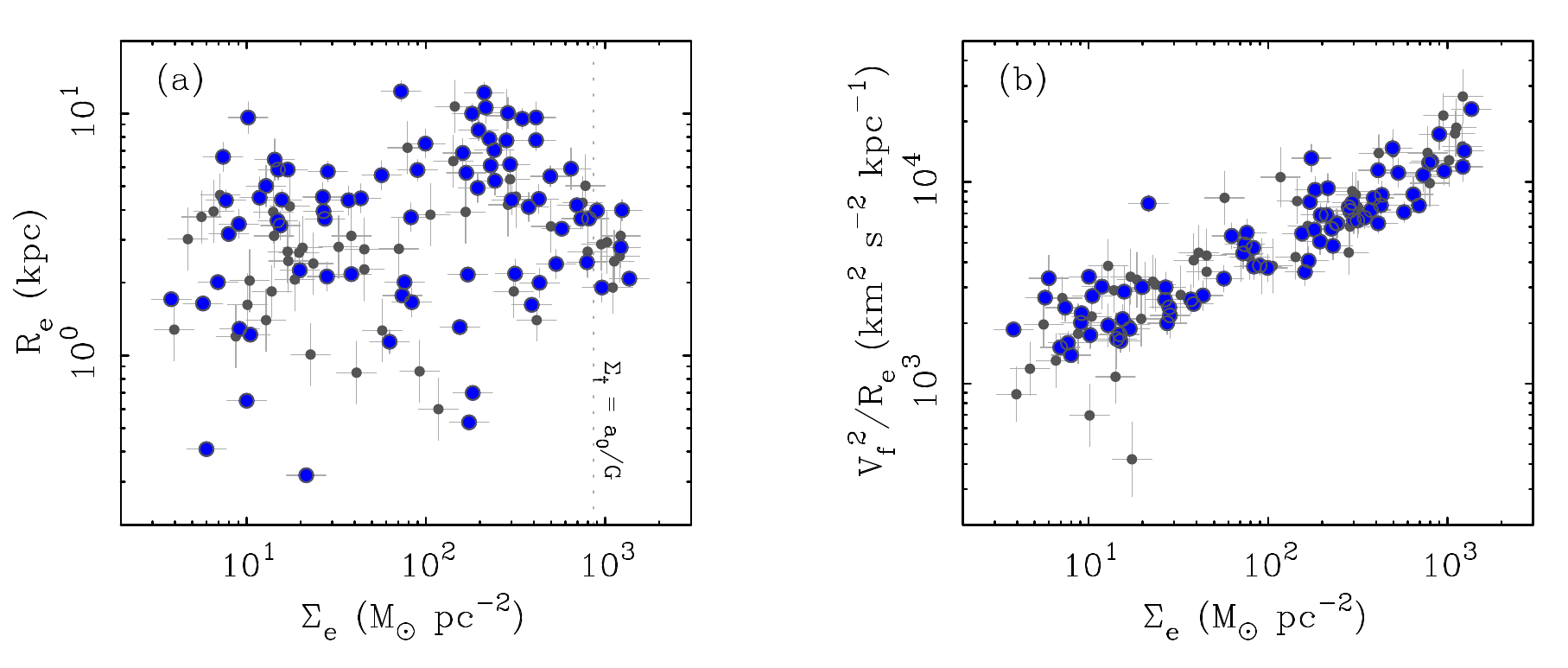}
\caption{The effective stellar mass surface density of galaxies~\cite{SPARC} as functions of their (\textbf{a}) effective radii and (\textbf{b}) characteristic 
accelerations. A galaxy must have a rotation curve that is extended enough to measure \Vf\ \cite{SPARCTF} to appear in this diagram. 
Larger points are galaxies with distances that are accurate to better than 20\%; smaller points have less accurate distances.
Effective surface brightness is converted to a surface mass density assuming a [3.6] $\ML = 0.5\;\MLsun$.
Galaxies exist over a wide range in size and surface brightness, with~no particular correlation, up~to a practical maximum in each (\textbf{a}). 
In~contrast, there is a strong correlation between the characteristic acceleration and surface brightness (\textbf{b}), as~anticipated by MOND.
The scale $\Sigma_{\dagger} = \azero/G$ is noted as a dotted line in (\textbf{a}). 
\label{ReSeAcc}}
\end{figure}

\paragraph{\textit{Do the data corroborate the prediction of MOND?}} Yes.
This is apparent directly from the observed dynamical accelerations (Figure \ref{ReSeAcc}). These vary in direct correspondence
to the observed $3.6\,\mu$ surface brightness. Low surface brightness systems have low accelerations; high surface brightness
galaxies display high accelerations. There is a clear continuum from one end of the galaxy spectrum to the other.
This happens despite the enormous scatter in the size--surface brightness~plane.

\paragraph{\textit{Was the prediction made} a priori?} Yes. 
Figure~\ref{ReSeAcc} is purely empirical. It simply plots the data~\cite{SPARC}; there is no fitting of any sort.
The~correlation apparent in Figure~\ref{ReSeAcc} directly indicates
the connection between surface brightness and acceleration, which traces the dynamical surface density.
That this should happen was predicted \textit{a priori} by MOND at its~inception.

\paragraph{\textit{What does dark matter predict?}} In order to predict how surface density correlates with surface brightness, 
one needs to know both. Dark matter-only simulations provide  excellent predictions for what the density profiles of dark matter
halos should be~\cite{NFW}, but~are mute about the surface brightnesses of the galaxies they contain. Hydrodynamical simulations
obtain a variety of results for the distribution of baryons, and~there is no clear consensus about what this should be~\cite{Aquila}.
Consequently, \LCDM\ makes no clear prediction for this~observable. 

\begin{Property} 
Predicting Rotation Curves 
\label{pr:fits}
\end{Property}

In MOND, the~dynamics should follow from the observed mass distribution.
To perform this test, we need to calculate the Newtonian gravitational potential associated with the observed mass,
calculate the corresponding force in MOND, and~observe a tracer of that force. Rotation curves provide 
a test in which this ideal is nearly~achieved.

\begin{figure}[h]
\centering
\includegraphics[width=6 in]{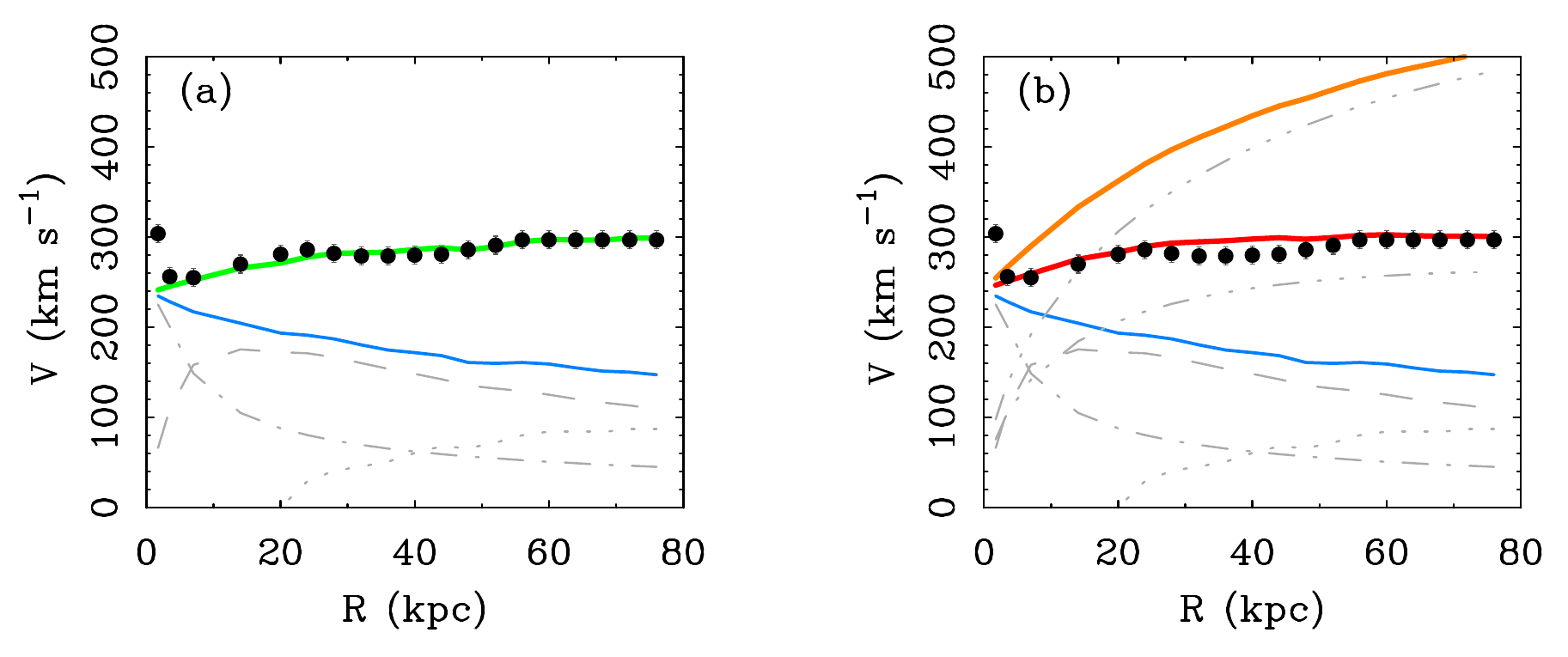}
\caption{Rotation curve and mass models for the giant spiral galaxy UGC 2885 ($M_* \approx 2 \times 10^{11}\,\Msun$) 
in (\textbf{a}) MOND and (\textbf{b}) \LCDM.
Mass models of the individual components are shown as gray lines: dotted for the gas,
dashed for the stellar disk (for $\ML = 0.5\;\MLsun$ at [3.6]), dash-dotted for the bulge ($\ML = 0.7\;\MLsun$), 
and dash-triple dotted for the dark matter halos in (\textbf{b}). 
A thin blue solid line shows the sum of baryonic components: this is the expected rotation without dark matter or MOND. 
A thick solid line shows the corresponding rotation that is \textbf{\textit{predicted}} in MOND (green line in \textbf{a})
and \LCDM\ (red and orange lines in \textbf{b}). In~the latter case, two approaches
are taken to predict the mass of the dark matter halo (see text). Assuming $m_d = 0.05$ \cite{MMW98} results in the red line
that performs almost as well as MOND. Using abundance matching~\cite{BBK} results in the orange line that overshoots the data.
\label{UGC2885}}
\end{figure}

It is possible to \textit{predict} rotation curves from the {baryonic} mass distribution of 
galaxies: {the atomic gas is traced by 21cm observations while the stellar mass} is well-traced by the near-IR light.
With~the adopted mass-to-light ratio, we convert the surface brightness profiles of galaxies observed by Spitzer~\cite{SM14Spitzer} 
into mass models~\cite{SPARC} that represent the gravitational potential of the stars. The~same has been done for the gas
in the course of obtaining 21cm rotation curves (see the many references in~\cite{SPARC}). These mass models are
representations of the Newtonian gravitational potential of stars and gas (e.g., Figure~\ref{UGC2885}).
These potentials add linearly and predict radial accelerations $g_N = -\partial \Phi/\partial R$ that must match the
centripetal acceleration to sustain circular~motion.

The observed stars and gas ($g_N$) fall short of explaining the centripetal acceleration indicated by rotation curves ($a = V^2/R$)---hence the need for dark matter or MOND. In~the former case, we simply attribute any excess to dark matter. 
In MOND, there is a mathematical relation between what we see and what we get~\cite{QUMOND}:
\begin{equation}
a = \nu(g_N/\azero) g_N
\label{eq:mondift}
\end{equation}
where $\nu(g_N/\azero)$ is an interpolation function that smoothly joins the high and low acceleration 
regimes~\cite{milgrom83a,MilgSand2008,M08}. 
This is not specified theoretically, but~is constrained empirically to be something very close to the 
so-called "simple" function~\cite{FBMW,Hees2016}. Here we adopt~\cite{M08}
\begin{equation}
\nu^{-1} = 1 - e^{-\sqrt{g_N/\azero}},
\label{eq:interp}
\end{equation}
which describes the data well~\cite{RAR}. 
Once we specify this function, we can use Equation~(\ref{eq:mondift}) to predict rotation curves from the baryonic mass~distribution.

Figure~\ref{UGC2885} shows an example rotation curve prediction.
First, the~Newtonian acceleration $g_N$ is estimated using the nominal mass-to-light ratios for the stellar components.
We then use Equation~(\ref{eq:mondift}) to obtain the MOND-predicted acceleration. This is shown as the green line ($V = \sqrt{aR}$)
in Figure~\ref{UGC2885}a. This provides a remarkably good match to the data for a hands-free prediction.
Only the first point is missed; this is because our nominal bulge mass-to-light ratio in a bit small for this galaxy---in a fit,
it~grows to $0.97\;\MLsun$ \cite{LiRAR}. This is within the range of expected variation. 
Moreover, this must be the case in either theory. It occurs in the high acceleration limit, so MOND gives no boost. 
Nor can we invoke a dark matter halo, as~the rotation curve declines steeply after the first point, just as the shape
of the bulge light distribution predicts, while the rotation curve of the dark matter halo must rise monotonically if it
is to fit the data further out. In~either case, we need a higher mass-to-light ratio for the~bulge.

\begin{figure}[h]
\centering
\includegraphics[width=6 in]{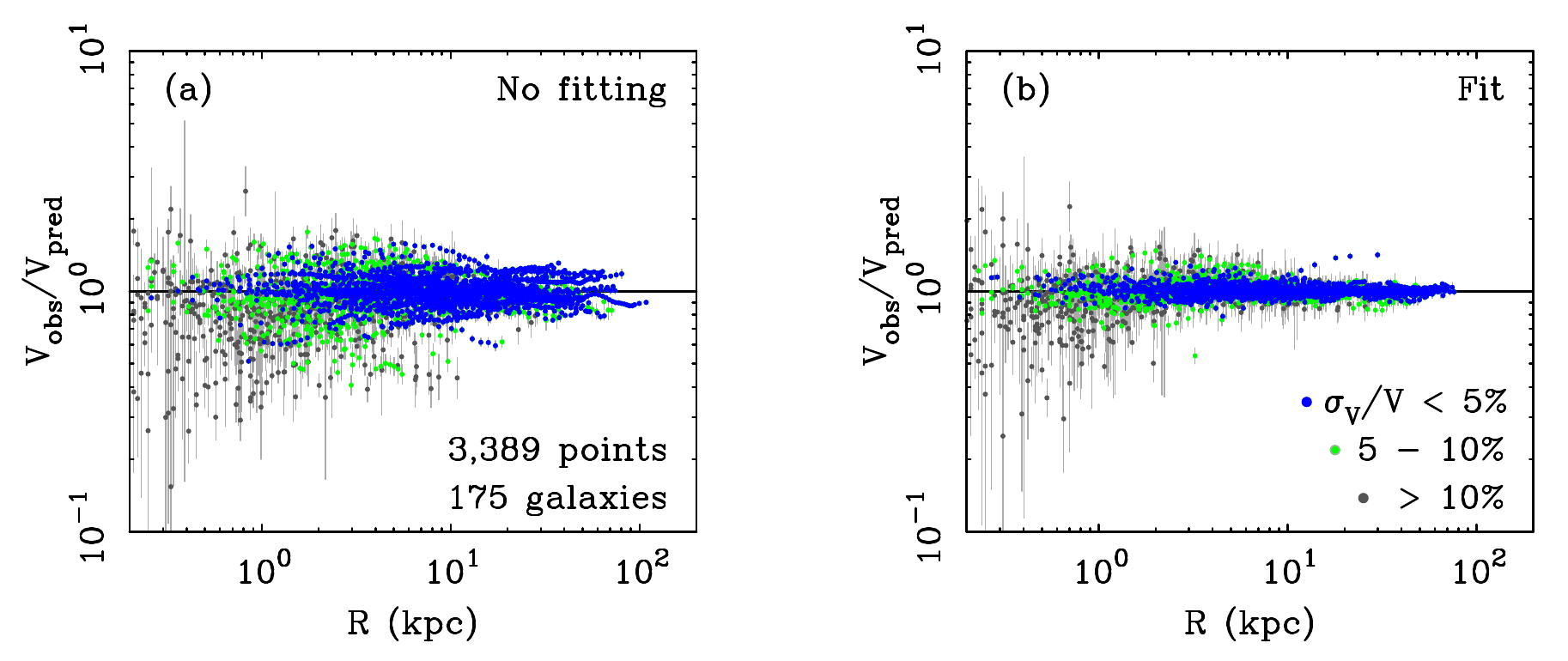}
\caption{Ratio of the observed velocity to that predicted by MOND, without~(\textbf{a}) and with (\textbf{b}) fitting.
All available data for galaxies from the SPARC database~\cite{SPARC} are shown. 
Each point represents one resolved datum along the rotation curves of these galaxies. 
Points are color coded by measurement accuracy, as~noted in the inset.
No fitting has been performed in (\textbf{a}): the same nominal mass-to-light ratio ([3.6] $\ML = 0.5\;\MLsun$ for the disk
and $0.7\;\MLsun$ for the bulge) has been adopted for all galaxies to \textbf{\textit{predict}} the velocity---i.e.,~the equivalent
of the green line in Figure~\ref{UGC2885}a for \textit{all} SPARC galaxies. This~procedure returns the correct velocity 
to within 0.15 dex for 90\% of the data. The~small scatter in (\textbf{a}) 
is further reduced (\textbf{b}) by fitting~\cite{LiRAR} for the optimal mass-to-light ratio of each galaxy (Figure \ref{MLSBpop}).
\label{mondpredacc}}
\end{figure}

The successful MOND prediction of the rotation curve of UGC 2885 seen in Figure~\ref{UGC2885} is not a fluke; it is the general rule.
Figure~\ref{mondpredacc} shows the residuals of MOND-predicted rotation curves for 175 galaxies for which all necessary, credible
data (a Spitzer map of the stellar mass, an~HI map of the atomic gas mass, and~a rotation curve) are available.
No fitting has been performed in Figure~\ref{mondpredacc}a, which simply plots the ratio of the observed velocity
to that predicted by MOND for the nominal mass-to-light ratio. The~same \ML\ has been assumed for all galaxies:
what you see is what you get. That~it was possible to effectively predicted rotation curves with near-IR surface photometry
was also noted by~\cite{SV1998}. The~same holds in galaxies where atomic gas is the dominant form of 
baryonic mass~\cite{Sanders2019}, as~the stellar mass-to-light ratio matters little for such~galaxies. 

The result of fitting the data~\cite{LiRAR} is shown in Figure~\ref{mondpredacc}b.
The scatter declines as expected, albeit by a modest factor: the raw prediction with a constant mass-to-light ratio for all galaxies
in~Figure~\ref{mondpredacc}a is already pretty good. The~reduction in scatter here manifests in an increased scatter in the
stellar mass-to-light ratio (see below). This \textit{must} happen; a constant \ML\ makes for a nice, hands-free assumption, but~there must be some intrinsic scatter in this quantity. It turns out that the scatter so induced is about that expected from variations
in the star formation histories of galaxies~\cite{meidt}. There is the expected amount of variation in the mass-to-light ratio, leaving
little room for intrinsic scatter in the underlying~relation.

A subtle point worth noting is that the deviations seen at small radii in Figure~\ref{mondpredacc} velocity skew preferentially to 
$V_{obs} < V_{pred}$. This is the effect that is expected from the combination of observational resolution ("beam smearing") and
asymmetric drift (non-circular motion). It is hard to measure the velocity accurately as small radii where the gradient of the rotation curve
is large so that different velocities contribute within the first beam; the result is often an underestimate of the true rotation speed. 
It is also the case that non-circular motions sometimes make up a large fraction of the kinetic energy at small radii so that the measured velocity 
sometimes falls short of the desired circular velocity of the gravitational potential.
These effects both result in a systematic skew in the sense observed, particularly in the lowest quality data (the grey points in
Figure~\ref{mondpredacc}). 
 
\begin{figure}[h]
\centering
\includegraphics[width=6 in]{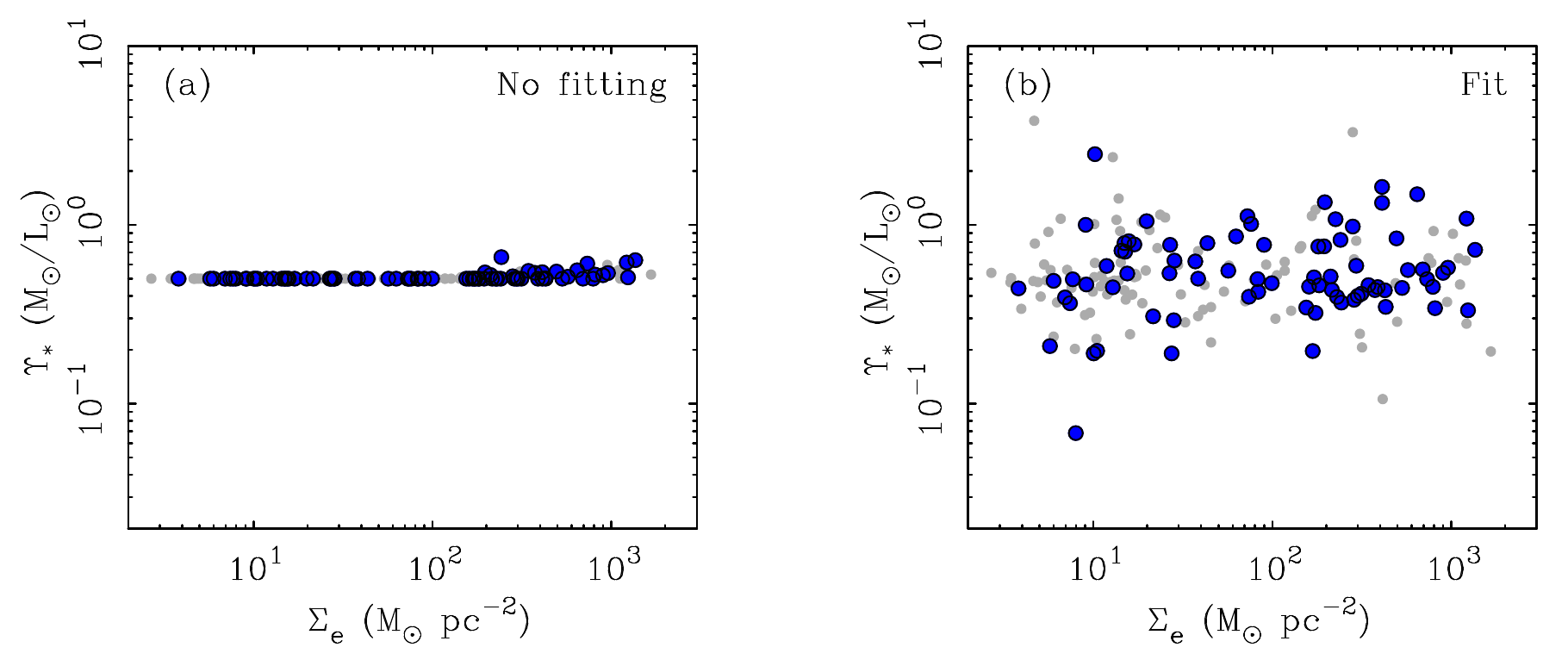}
\caption{The effective surface densities of SPARC galaxies and their stellar mass-to-light ratios as obtained  
(\textbf{a}) from assuming constant $\ML = 0.5\;\MLsun$ at [3.6] for stellar disks and $\ML = 0.7\;\MLsun$ for bulges,
and (\textbf{b}) by fitting rotation curves~\cite{LiRAR}. Large blue points are galaxies with distances known to better than 20\%;
smaller grey points are galaxies with less accurate distances. The~absence of scatter in (\textbf{a}) is anathema to stellar populations;
there must be some intrinsic scatter in this quantity from variations in the star formation history from galaxy to galaxy.
The scatter seen in (\textbf{b}) is consistent with that expected from intrinsic scatter in \ML\ \cite{meidt,SM14,SML19}
and observational uncertainties~\cite{RAR,OneLaw,LiRAR}. 
\label{MLSBpop}}
\end{figure}

\paragraph{\textit{Do the data corroborate the prediction of MOND?}} Yes. The~predictive ability of MOND 
is as good as can be expected given the fundamental limitation of converting the observed starlight
into the corresponding stellar mass. One can reduce the scatter in Figure~\ref{mondpredacc}a by treating \ML\ as an adjustable parameter.
The efficacy of this procedure is apparent in Figure~\ref{mondpredacc}b and 
has been demonstrated many times before~\cite{BBS,S1996,dBM98,SMmond,swatersmond,Gentile2011,LivRev}.
There~are, of~course, exceptions: galaxies that do not fit in detail. For~example, NGC 2841 was long considered
problematic~\cite{N2841}, but~a good fit falls out of a Bayesian analysis~\cite{LiRAR}. Still, problematic cases persist (e.g., NGC 2915).
There are always cases like this in astronomy; it would be suspicious if all the data could be fit without some outliers. The~failure rate
increases as data quality declines, as~expected. We~should not lose sight of the forest for the occasional outlying~tree.

\paragraph{\textit{Was the prediction made} a priori?} That it should be possible to predict rotation curves from the observed
mass distribution of galaxies was predicted \textit{a priori}. The~extent to which this is possible for any individual galaxy is limited
by astrophysical uncertainties in how well we can measure the mass distribution; in~particular, the unavoidable uncertainty in \ML. 
That it is possible to come as close as illustrated by Figure~\ref{mondpredacc} is a remarkable~accomplishment.
 
\paragraph{\textit{What does dark matter predict?}} CDM\ makes clear predictions for the rotation curves of dark matter halos~\cite{NFW}. 
Predictions for the observable properties of galaxies are model-dependent. Many different models are possible;
Figure~\ref{UGC2885} illustrates two~possibilities. 

In order to predict the rotation curve of a specific galaxy, we need a mechanism to specify the dark matter halo within which it resides.
Perhaps the most obvious mechanism at present is offered by the stellar mass--halo mass relation obtained from abundance matching~\cite{BBK}.
A massive galaxy like UGC 2885 with $M_* \approx 2 \times 10^{11}\;\Msun$ should reside in a halo of mass $M_{200} \approx 5 \times 10^{13}\,\Msun$
(see Figure~6 of~\cite{BBK}). Together with the halo mass--concentration relation~\cite{DM2014}, this predicts the expected rotation attributable
to the dark matter halo. Adding this in quadrature with the baryonic component results in the orange line depicted in Figure~\ref{UGC2885}b. 
This grossly over-predicts the observed~rotation. 

If one did this exercise twenty years ago (I did), then the stellar mass--halo mass relation from abundance matching was not yet available.
The common approach then was to assume a constant disk to halo ratio around $m_d \approx 0.05$ \cite{MMW98}. Adopting this, we predict
$M_{200} \approx 4 \times 10^{12}\;\Msun$. This comes much closer to matching the observed rotation, performing almost as well as MOND
in predicting the rotation~curve. 

The good performance of assuming a disk fraction $m_d \approx 0.05$ is certainly a fluke.
We could just as easily have assumed $m_d = 0.1$ or 0.025 (both values considered by~\cite{MMW98}), and~we would again mispredict the rotation curve.
There is no reason to expect, \textit{a prior}, that this particular galaxy should have this particular disk fraction. 
We can fit the data to infer $m_d$, but~we cannot predict the rotation curve. If~we use one galaxy to fix $m_d$, we then get incorrect results
for other galaxies: $m_d$ must vary with mass~\cite{M10}.

The stellar mass--halo mass relation of abundance matching is now an essential element of the \LCDM\ paradigm~\cite{WT18}, 
so the discrepancy of the optimal disk fraction from this relation cannot be ignored.
It is tempting to conclude that this particular galaxy happens to be an outlier in the scatter about the mean $M_*$--$M_{200}$ relation,
by chance having a small total mass for its observed stellar mass. This is equivalent to suggesting that it has an abnormally low velocity
(the green line in Figure~\ref{UGC2885} rather than the expected orange line). This in turn predicts that it should sit far off of the Tully--Fisher 
relation defined by other galaxies of the same stellar mass. It does not. Indeed, in~general, there is too little scatter in the BTFR to accommodate
that expected in the stellar mass--halo mass~relation.

There exist many other possibilities in the context of \LCDM\ that are not considered here.
Indeed, we have ignored processes that must be relevant, like adiabatic compression of the halo~\cite{adiabat}, and~any form of stellar feedback
(though this is usually said not to be important in galaxies of this high mass). It seems common to imagine that the solution lies in getting the
combination of these effects right, but~really this makes the problem worse, not better: there is no unique way to predict rotation curves with \LCDM. 
A huge number of models are possible; many are plausible. Nature appears to have declined to implement any reasonable \LCDM\ model.
The best we can hope to do is very precisely mimic the behavior of MOND, reproducing after the fact the phenomenology it correctly predicted in~advance.

\begin{Property} 
Stellar Population Mass-to-Light Ratios
\label{pr:pops}
\end{Property}

The stellar mass-to-light ratio is the only physical parameter available to MOND fits.
A~considerable amount is known about stellar populations, so these provide an independent check. 
If~MOND is simply a strange fitting function, there is no need for its fitting parameter to return plausible mass-to-light ratios.
If instead there is something to it, then the fitted values of \ML\ should make sense in terms of stellar~populations.

Figure~\ref{MLSBpop}a shows the stellar mass-to-light ratios for SPARC galaxies as assumed throughout this work
([3.6] $\ML = 0.5\;\MLsun$ for stellar disks and $\ML = 0.7\;\MLsun$ for bulges).
The~luminosity-weighted mass-to-light ratio is shown, so the slight variation seen for a few points is from differences in the bulge fraction. 
Most galaxies appear as beads on a string. This morphology is anathema to stellar populations, which must inevitably
suffer scatter from variations in the star formation history, the~metallicity distribution of the stars, and~differences in the 
IMF (Initial Mass Function: the distribution of masses with which stars form). 
In short, the~absence of scatter in Figure~\ref{MLSBpop}a is unphysical. 
This suffices only as a first estimate, but~there must be some intrinsic scatter in \ML. 

In Figure~\ref{mondpredacc} we predicted rotation curves using the \ML\ shown in Figure~\ref{MLSBpop}a.
Figure~\ref{MLSBpop}b shows the stellar mass-to-light ratios obtained from rotation curve fits~\cite{LiRAR}.
This is what is required to eliminate nearly all the scatter in Figure~\ref{mondpredacc}, transferring it from deviations in the predicted velocity
to the scatter that appears here. The~amount of scatter required to make rotation curve fits could have been arbitrarily large.
Instead, it is rather modest. Indeed, 
the scatter in \ML\ seen in Figure~\ref{MLSBpop}b is consistent with that expected from the combination of observational errors  
and intrinsic scatter in stellar population \ML\ ($\sim$$0.11$ dex at [3.6] \cite{meidt}) stemming simply from variations in the star formation
history. There is little room for other plausible sources of variation, like galaxy-to-galaxy differences in the average IMF. 
Given that some intrinsic scatter in \ML\ is inevitable, it is hard to imagine a more favorable~outcome.

\paragraph{\textit{Do the data corroborate the prediction of MOND?}} Yes. The~stellar mass-to-light ratios of 
MOND fits (Figure \ref{MLSBpop}; \cite{SMmond,LivRev,LiRAR}) are in excellent accord with the expectations of stellar population 
models~\cite{BdJ,portinari,meidt,SM14,MS14,SML19}. The~amplitude of \ML\ is consistent with what is expected for a Kroupa or
Chabrier IMF, which are practically indistinguishable. Heavier or lighter IMFs are disfavored. 
The scatter increases from red to blue bandpasses, as~expected, and~the expected color-\ML\ relations are also
recovered~\cite{LivRev,MS15}.

\paragraph{\textit{Was the prediction made} a priori?} No, and~it cannot be. 
The test here is whether the mass-to-light ratios required in MOND fits are consistent with the astrophysical expectations 
of stellar population models. To~a remarkable extent, they~are.

\paragraph{\textit{What does dark matter predict?}} I am not aware of any mechanism by which a similar test could be made in \LCDM\ 
{as the stellar mass-to-light ratio does not play an equivalent role in determining the dynamics that specify the rotation curve.
In MOND, \ML\ is uniquely specified, within~the uncertainties, while in dark matter models there is an unavoidable degeneracy between
dark and luminous mass~\cite{K87}, precluding a unique~test.}

\begin{Property} 
The Correspondence of Features
\label{pr:renzo}
\end{Property}

An important aspect of galaxy dynamics is the observed correspondence between features observed in 
the baryonic mass distribution and those seen in rotation curves. 
The "bumps and wiggles" in one are reflected in the other. 
This is known among experts as "Renzo's Rule" \cite{Sancisi04}.

It was recognized early on~\cite{vAS1986} that the observed correspondence of bumps and wiggles 
implied that stars were the dominant mass component at small radii.
The correspondence exists because the stellar mass dominates the gravitational potential, so features in the stellar distribution
are necessarily reflected in the rotation curve. This situation is generally known as \textit{maximum disk}: 
the stellar mass is close to the maximum allowed by the rotation curve~\cite{Sellwood99,palunas,Freeman10}.  

In HSB galaxies, the~maximum disk mass is generally comparable to or slightly higher than what is expected for 
stellar populations (0.7 vs.\ $0.5\;\MLsun$ at [3.6] for disks and 0.8 vs.\ 0.7 for bulges, 
with substantial individual variation~\cite{StarkmanMaxDisk}). In~these situations, the~correspondence between photometric 
and kinematic features is natural: the stars dominate the gravitational potential at small radii. The~bulge of UGC 2885
in Figure~\ref{UGC2885} is one example. There are many others~\cite{Sellwood99,palunas,StarkmanMaxDisk,IAU_review}.

The situation is different in LSB galaxies. Since the stellar mass is spread over a greater radius, the~contribution of
the stars to the total velocity is reduced simply because $V_*^2 \propto M_*/r$. For~the masses expected for stellar populations,
LSB galaxies are far removed from being maximal. One might choose to favor the maximum disk mass-to-light ratio over the stellar population
expectation, but~this would violate the constraints discussed above: LSB galaxies cannot be maximal and also fall on the BTFR.
Imposing maximum disk mass-to-light ratios would induce surface-brightness correlated scatter in Figure~\ref{fig:BTFR} that 
is not present in the raw~data.

\begin{figure}[h]
\centering
\includegraphics[width=6 in]{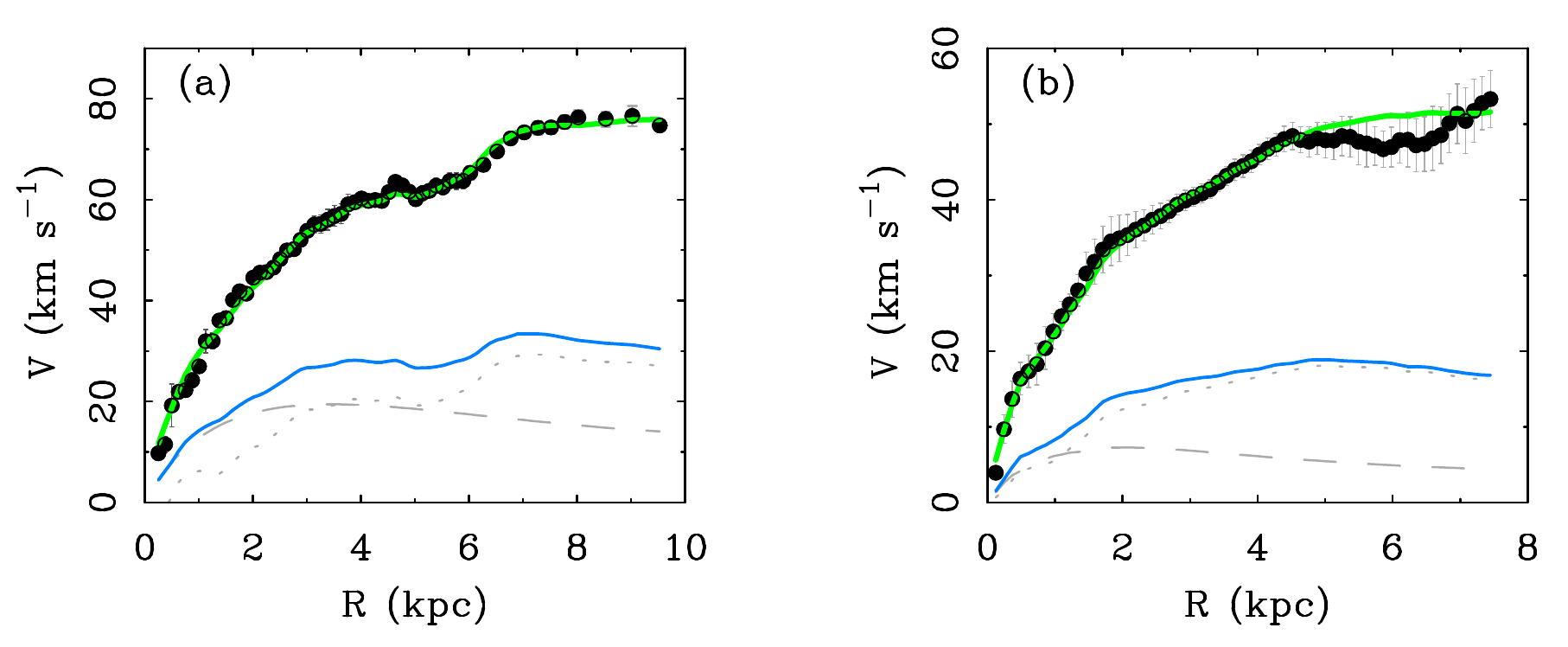}
\caption{Rotation curves and mass models for the dwarf galaxies (\textbf{a}) NGC 1560~\cite{BBS,gentileN1560,LivRev} 
and (\textbf{b}) DDO 154~\cite{THINGS,Gentile2011,LiRAR}. 
The MOND fits (green lines) necessarily follow the detailed shape of the features seen in the baryonic mass distributions (light blue lines). 
Gas (dotted lines) dominates the mass budget in these low surface density galaxies; the stars (dashed lines) and their
mass-to-light ratio have little leverage on the fit. 
\label{DDO154NGC1560}}
\end{figure}

Thin, dynamically cold stellar disks can support features like spiral arms while quasi-spherical, dynamically hot 
dark matter halos cannot~\cite{BT}. One therefore expects the correspondence between features to dissipate as 
surface brightness decreases and the dark matter halo comes to dominate. 
Nevertheless, the~correspondence of features persists 
(Figure \ref{DDO154NGC1560}; \cite{Sancisi04,VdB99,Swaters09,Swaters12,LelliVRgrad}).

Figure~\ref{DDO154NGC1560} shows two examples that illustrate the correspondence
of photometric and kinematic features in low surface density galaxies. 
NGC 1560 has a prominent dip from 5--6 kpc in both the baryonic and total rotation curve~\cite{BBS,gentileN1560}.
DDO 154 has a more subtle correspondence between the two, with~kinks around 0.5, 2, and~5 kpc~\cite{THINGS}.
These are not happenstance; this is the general rule~\cite{Sancisi04,LiRAR}: 
details like this are subsumed in the residuals for all the galaxies in Figure~\ref{mondpredacc}. 

The majority of the baryonic mass in the galaxies in Figure~\ref{DDO154NGC1560} is in the form of gas, not stars. 
Consequently, there is no leverage to fit the data by adjusting the stellar mass-to-light ratio: the shape of the rotation curve
follows directly from the observed distribution of gas. As~emphasized by~\cite{Sanders2019}, galaxies like these provide very nearly 
a direct prediction without any fitting. However, there are nuisance parameters that need to be considered~\cite{LiRAR}.
The distance and inclination is measured independently for each galaxy, but~of course these are not known perfectly well.
These influence the baryonic mass ($M \propto d^2$) and rotation speed [through $\sin(i)$]. The~case of DDO 154 provides
a good illustration of both effects. Distance estimates to DDO 154 
range\footnote{The NASA/IPAC Extragalactic Database (NED) is operated by the Jet Propulsion Laboratory, 
California Institute of Technology, under~contract with the National Aeronautics and Space Administration.} from 3 to 6 Mpc. 
The~formally most accurate measurement is $d = 4.04 \pm 0.08$ Mpc~\cite{EDD,EDD2}. If~we hold the distance
fixed at 4.04~Mpc, then~the shape of the rotation curve is the same but the amplitude slightly overshoots the data
(see~Figure~S5 of~\cite{RKKYPRX}). Distances are never known perfectly; treating it as a nuisance parameter in a Bayesian
fit with a prior that matches the measurement uncertainty leads to $d = 3.87 \pm 0.16$ Mpc~\cite{LiRAR}. This~small reduction
in the distance is the difference between overshooting the data and the excellent fit seen in Figure~\ref{DDO154NGC1560}.
Similarly, the~inclination is not perfectly well known. In~the case of DDO 154, it~becomes particularly uncertain at large radii 
(see Figure~81 of~\cite{THINGS}) where the shape of the rotation curve becomes dodgy. The~slight mismatch in the shape
of the MOND fit in the outer fringes of DDO 154 is a good thing: it cannot be fooled into tracing unphysical variations~\cite{dBM98}. 

A good theory should not only fit the data, it should also fail to fit data that are incorrect.
To~test this, an~outright mistake in the baryon distribution was intentionally introduced by~\cite{dBM98}. 
An~acceptable MOND fit to these incorrect data could not be found: confronted with a situation in which it \textit{should} fail, it did so.
In contrast, there is substantially more freedom in fits with dark matter halos: one could happily fit the data 
without noticing that the baryonic distribution was wrong, much less notice a detail like a slight issue with the~distance. 

The experiment of using an incorrect baryon distribution has been unintentionally replicated by~\cite{RKKYPRX} in the case of D631-7 
(the first example in their Figure S5). D631-7 is a gas rich galaxy similar to those in Figure~\ref{DDO154NGC1560}. 
However, only a total gas mass is available; the detailed gas distribution is not. 
For inclusion in the SPARC database~\cite{SPARC}, a~scaling relation between gas mass and radius was applied to make a crude 
estimate of the gas distribution~\cite{SPARC,OneLaw}. This estimate is certainly wrong in detail, and~
indeed, the~MOND fit that~\cite{RKKYPRX} obtain using it is a poor match to the data---as it should be in such a circumstance. 
In contrast, the~fit~\cite{RKKYPRX} make with dark matter shows no indication of a problem. 
There is sufficient freedom in fits with dark matter halos
to absorb even gross errors in the input data; they are incapable of failing when they should~\cite{dBM98}.

\paragraph{\textit{Do the data corroborate the prediction of MOND?}} Yes. In~MOND, the~detailed shape of the rotation curve
must follow from the observed distribution of mass. This is what is~observed.

\paragraph{\textit{Was the prediction made} a priori?} Yes and no. That this should be generally be the case was
anticipated in the original papers~\cite{milgrom83b}. For~specific galaxies, this prediction must be made 
on a case by case basis. Figure~\ref{mondpredacc} illustrates how well rotation curves can be~predicted.

\paragraph{\textit{What does dark matter predict?}}  The conventional expectation is that dark matter halos should not support
the same features that are seen in the luminous disk. Dynamically hot dark matter halos that dominate the mass budget should
not be affected by the small minority of mass in the disks of LSB galaxies, and~are not able to sustain similar features on their own~\cite{BT}.
Any one case might be dismissed as a happenstance of some non-equilibrium event, but~the specific cases illustrated in
Figure~\ref{DDO154NGC1560} are not the exception; they are examples of the general rule (Figure \ref{mondpredacc}).
The widespread correspondence between features in the baryonic mass profiles and the kinematics of LSB galaxies is contradictory to
any flavor of dark matter that does not interact with baryons by some mechanism more direct than~gravity.

\subsection{Disk~Stability}

There are many indications of mass discrepancies in extragalactic astronomy and 
cosmology~\mbox{\cite{FG79,MdB98a,SMmond,LivRev,CJP}}. One of the early indications was disk stability. Left to themselves, spiral
disks that are not embedded in dark matter halos are subject to a violent bar instability~\cite{OP73}.
Maintaining thin, stable, dynamically cold spiral disks for the better part of a Hubble time seems to require 
some assistance~\cite{Sellwood14,Sellwood16,SSZM33}. A~simple way to think of this is a competition between disk self-gravity,
which~drives instabilities like bars and spiral arms, and~the gravity of a dynamically hot dark matter halo, which tends to suppress
these instabilities. Explaining the observed morphologies of spiral disks requires some of~both.

\begin{Property} 
The Freeman Limit
\label{pr:freeman}
\end{Property}

The highest surface brightness galaxies have the most disk self-gravity, so are most subject to self-destructive instabilities.
These HSB galaxies are at the Freeman limit, which is a generalization of Freeman's Law~\cite{F70}. 
LSB galaxies exist in great numbers~\cite{Disney,AllenShu,MBS95,CrossDriver}; what was called Freeman's Law is 
not a constancy of surface brightness for all galaxies, but~an upper limit on surface brightness that disk galaxies 
do not exceed~\cite{AllenShu,MBS95}. 

A first investigation of disk stability in MOND was discussed in~\cite{MOND_Freeman}, 
and numerical simulations have been conducted by~\cite{bradastab,TiretCombes_stab,JH2014,Ingo,SS16,Indranil18}. 
The basic result is that MOND stabilizes galaxy disks without a dark matter halo.
There are two essential predictions that appear already in the first work~\cite{MOND_Freeman} and persist in 
the numerical simulations: disk galaxies can only exist in the MOND regime,
and the amount of stability predicted for LSB galaxies differs from that expected with dark matter~halos.

\paragraph{\textit{Do the data corroborate the prediction of MOND?}} Yes.
Bare Newtonian disks should suffer the usual instability in the absence of dark matter, so are predicted not to exist.
This sets an upper limit to the surface brightness, as~stabilizing accelerations are only obtained for $a < \azero \approx G \Sigma_{\dagger}$.
The observed value of the Freeman limit corresponds well to $\Sigma_{\dagger}$ (Figure \ref{ReSeAcc}; \cite{LivRev}).

Note that the scale \azero\ appears in disk stability in a way that is different from its appearance in galaxy kinematics.
In kinematic relations like the BTR, it appears with Netwon's constant as the product $\azero G$.
In disk stability, it appears through the ratio with Newton's constant: $\Sigma_{\dagger} = \azero/G$.
Hence the scale \azero\ appears in galaxy data in distinct ways that are unique to~MOND.

\paragraph{\textit{Was the prediction made} a priori?} No and yes. 
The Freeman surface brightness was known before MOND was invented, and~before the first investigation of disk stability therein.
However, it was correctly anticipated~\cite{MOND_Freeman} that the Freeman surface brightness was a limit rather than a universal value
at a time when most of the community interpreted it to be the~latter.

\paragraph{\textit{What does dark matter predict?}} Disk stability in CDM\ depends, crudely speaking, on~the disk-to-halo ratio. 
If this is too large, the~disk becomes unstable. The~dense, cuspy halos that emerge from numerical simulations are capable of 
stabilizing disks of considerably higher surface density than the Freeman limit~\cite{cusp_stab}; 
the scale $\Sigma_{\dagger}$ had to be inserted into models by hand~\cite{DSS97,MdB98a}. 
There is no reason that the threshold for stability should be that predicted by MOND, as~observed. 

\begin{Property} 
Vertical Velocity Dispersions
\label{pr:vvd}
\end{Property}

\begin{quote}
"An analog of the Oort discrepancy should exist in all galaxies and become more severe with increasing [radius] in a predictable way."\\
---M.\ Milgrom~\cite{milgrom83b}.
\end{quote}

The vertical velocity dispersions of disk galaxies are related to their stability. Disks are dynamically cold, in~the sense that
$\sigma_z \ll V_c$ \cite{OhThings,diskmass}. Cold disks are especially subject to instabilities~\mbox{\cite{OhThings,diskmass}}, which 
was an important consideration driving early work~\cite{OP73} and remains an important consideration today. 
A related property is the Oort discrepancy; i.e.,~the excess vertical velocity dispersion over that which can be explained by the 
Newtonian restoring force to the stellar~disk.

Conventionally, the~Oort discrepancy should be modest. Near~the disk, the~stars dominate the mass budget and provide the lion's
share of the restoring force. It is only as one looks to high vertical distances from the center of the plane that one begins to notice
the contribution of the quasi-spherical dark matter~halo.

In MOND, the~amplitude of the discrepancy depends on the acceleration. In~high acceleration regimes, there should be no discrepancy.
The discrepancy should appear around \azero, and~grow larger as accelerations decrease. In~HSB galaxies, the~severity of the Oort
discrepancy should increase with radius because acceleration decreases with radius (Figure \ref{VRSB}b). Interpreted conventionally, one would infer
a dark matter halo that is very squashed near the disk plane, or~even a disk of dark matter, transitioning to a more spherical potential
farther out. The~quantitative details of how this occurs may be theory-specific: not all theories~\cite{AQUAL,QUMOND,milgrom94,milgrom99}
that follow the basic tenets of MOND~\cite{mondtenets} need necessarily be identical in this~regard.

\paragraph{\textit{Do the data corroborate the prediction of MOND?}} For this test, the~results are mixed. 
In general, the~\textit{shape} of the predicted velocity dispersion profile is often correct, but~the \textit{amplitude} is
frequently over-predicted. This makes no sense in either MOND or dark matter. MOND should get both right. 
In dark matter, the~shape of $\sigma_z(r)$ should follow the prediction of Newton~\cite{Das2020}, not MOND. 
A similar conundrum arises for clusters of galaxies~\mbox{\cite{sanders2003,sanders2007,angusbuote,CJP,YongCLASH}}.

In the Milky Way, the~rotation curve is well described by MOND, which successfully predicted its outer slope~\cite{M19}. 
However, the~vertical velocities are over-predicted~\cite{M16} by $\sim$$15\%$ \cite{Lisanti2018}. This~is about a $2\sigma$ discrepancy,
so dark matter is favored by the vertical velocity data, provided that we spot it MOND-like behavior in the radial direction.
It is not obvious that this make sense in principle, and~it leads to a puzzle in practice. The~local dark matter density inferred
from the rotation curve is $\sim$$0.007\;\Msun\,\mathrm{pc}^{-3}$ \cite{MW2018} while that from vertical motions implies twice as
much: $\sim$$0.014\;\Msun\,\mathrm{pc}^{-3}$ \cite{Beinayme}. This implies a squashed halo~\cite{Beinayme}, but~this is
contrary to the findings of~\cite{Lisanti2018} for which a spherical halo is a reasonable~fit.

In external galaxies, we encounter a similar problem. This challenging observation has been undertaken by the 
DiskMass project~\cite{diskmass_orig}, with~the result that disks are not merely cold dynamically, but~downright frigid.
Using conventional dynamics, the~observed vertical velocity dispersions imply stellar mass-to-light ratios 
that are a factor of $\sim$$2$ \cite{diskmass} or more~\cite{diskmass_swaters}
lower than expected for stellar populations~\cite{MS14}. This is equivalent to removing all stars of mass $<$$1.1\;\Msun$ from a Kroupa IMF.
This~is not a viable solution, as~it implies that the sun and lower mass stars that are numerically common locally 
do not exist in other galaxies.
The problem gets worse in MOND, which predicts larger velocity dispersions~\cite{AngusDiskMassMOND}.
However, the~shape of the radial variation $\sigma_z(r)$ is well-predicted~\cite{diskmass_milgrom}; the problem is a small
offset between the observed and predicted dispersion that remains roughly constant as $\sigma_z(r)$ varies by a large factor.
Since the result makes little sense in the conventional context~\cite{AngusDiskMassWrong,Aniyan2016}, it is not surprising that it does not
work for MOND either. As~in the Milky Way, we are confronted with a situation that is problematic for both~paradigms.

There are some qualitative suggestions of MOND-like behavior in this context. The~velocity dispersions of the gas in the
outer regions of galaxies are consistently higher than can be sustained by the restoring force of the Newtonian disk~\cite{NGC2915}. 
This leads to the inference of highly flattened dark matter halos~\cite{Olling96b} or dark disk components~\cite{Das2020} distinct
from quasi-spherical halos, or~some non-gravitational effect. That the velocity dispersion in these
low density regions are higher than expected conventionally is the qualitative signature of MOND. By~the same token, 
there exist ultrathin disk galaxies~\cite{ultrathin1,ultrathin2,ultrathin3} that are difficult to sustain conventionally with the weak restoring 
of their low surface density stellar disks: they should be much thicker than observed. 
That these galaxies are thin follows naturally from the enhanced restoring force provided by MOND (see Figure~9 of~\cite{MdB98b}).

There is no clear conclusion that can be drawn from the vertical velocity dispersion data at this time. 
There are a number of qualitative indications of MOND-like behavior, especially in low surface density systems where its effects should
be pronounced. However, its quantitative predictions persistently over-predict vertical velocities in the best observed systems, albeit
by a small~amount. 

\paragraph{\textit{Was the prediction made} a priori?} Yes. It remains the irrevocable prediction that the vertical velocity dispersion
should follow from the observed distribution of baryonic mass. However, the~details of the quantitative prediction may depend on
whether MOND is a modification of gravity~\cite{AQUAL,QUMOND} or inertia~\cite{milgrom94,milgrom99}.

\paragraph{\textit{What does dark matter predict?}} The prediction of conventional dynamics with dark matter depends on the 
detailed distribution of both dark and luminous mass. The~former is not observed, so we are free to assign to each dark matter halo 
whatever degree of flattening is required to fit the data. This is not as satisfactory as an \textit{a priori} prediction, but~it
should be possible to predict the distribution of halo shapes~\cite{DM2014,haloshape_dmo,haloshape_hydro} to make a statistical test. 
I am not aware of a conclusive observational test of this type. 

\begin{Property} 
Spiral Structure in LSB Galaxies
\label{pr:morph}
\end{Property}

\begin{quote}
"In LSB disks, it is conceivable that the minimum disk mass required to generate spiral arms might exceed the 
maximum disk mass allowed by the rotation~curve." \\
---S.\ McGaugh~\cite{MdB98b}
\end{quote}

For disk galaxies near the Freeman limit, the~stability provided by a dark matter halo is about the same as that provided by MOND. 
However, the~two theories diverge to lower accelerations. In~order to explain the amplitude of the rotation curve, the~disk-to-halo
ratio must steadily decline as the surface brightness declines: LSB galaxies are dark matter dominated. Consequently, they should
be very stable~\cite{MMdB}. In~contrast, the~stability provided by MOND does not continue to increase indefinitely in the regime of 
very low accelerations, instead saturating after a mild increase~\cite{MOND_Freeman,bradastab}.

The difference in the predicted stability properties of LSB disk galaxies leads to a difference in the expected morphology.
Dark matter halos over-stabilize low surface density disks, suppressing the development of 
bars and spiral arms~\cite{MMdB,Indranil18}. In~contrast, MOND predicts a more similar development of such features
in high and low surface brightness disks, with~numerical simulations showing remarkably realistic morphologies~\cite{TiretCombes_stab}.

Though dark matter halos were originally invoked to stabilize disks~\cite{OP73}, it was also recognized early that the disk-to-halo
ratio should not be too low, or~it would over-stabilize disks and suppress the observed spiral modes. This marginal stability condition
places a lower limit on the masses of the stellar disks~\cite{spiralmodes}. 
This minimum disk is not far removed from maximum disk for HSB galaxies.
In~contrast, LSB galaxies are well below maximum disk: for the stellar masses expected from population synthesis,
stars contribute little to the gravitational potential, even at small radii~\cite{M05,StarkmanMaxDisk}. 

\begin{figure}[h]
\centering
\includegraphics[width=6 in]{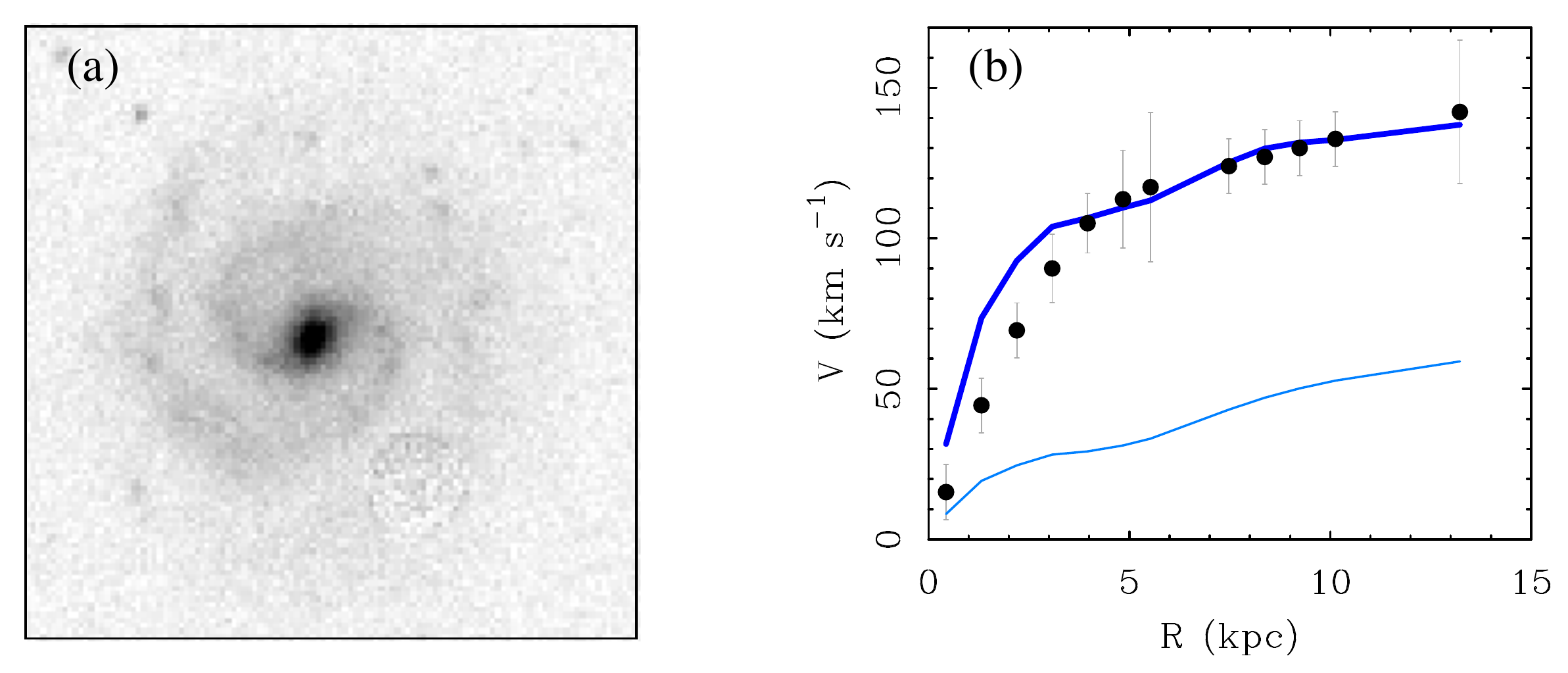}
\caption{The low surface brightness galaxy F568-1, as~seen (\textbf{a}) in the $V$-band~\cite{LSB2011}, 
together with (\textbf{b}) its rotation curve and mass model~\cite{SPARC}. The~lower blue line is the mass model for the
nominal stellar disk with $M_* \approx 3 \times 10^9\,\Msun$, which has a [3.6] $\ML = 0.5\,\;\MLsun$ and 
corresponding $V$-band $\ML = 1.4\;\MLsun$. 
The disk-to-halo ratio is small, so spiral structure should be suppressed~\cite{spiralmodes,MMdB}.
The upper blue line shows the disk rotation curve with the mass required 
to explain the observed spiral structure in (\textbf{a}) in the context of dark matter~\cite{Fuchs_ASS}. 
The stellar disk must be very heavy: 
$M_* \approx 42 \times 10^9\,\Msun$, with~$\ML = 6.7\,\;\MLsun$ in [3.6] and $20\;\MLsun$ in the $V$-band. 
This is well in excess of the mass expected for a stellar population. Indeed, the~stellar disk is so heavy that it 
leaves no room for dark matter and exceeds the observed rotation for $R < 3$ kpc.
Taken at face value, this poses a contradiction for any flavor of dark matter, as~predicted by~\cite{MdB98b}.
\label{Fuchs}}
\end{figure}

As a consequence, LSB galaxies should not exhibit bars or spiral structure if embedded in dominant dark matter halos~\cite{MMdB}. 
Though there are certainly differences in morphology between HSB and LSB galaxies, 
these are modest and there is no lack of examples of LSB galaxies with bars~\cite{PKdN18} and
spiral arms (Figure \ref{Fuchs} \cite{LSBmorphology,S2006,LSB2011}).
This observation is natural in MOND, as~there is ample disk self-gravity to drive the observed spiral structure,
and little dynamical friction to slow bars, which are observed to have higher pattern speeds~\cite{PKdN19} than expected
when dark matter dominates~\cite{DebSel,Weiner_bar}.

Figure~\ref{Fuchs} shows the example of the LSB galaxy F568-1. 
This galaxy is dim but large, with~a disk scale length $R_d \approx 5.2$ kpc~\cite{SPARC}, {somewhat larger than} 
that of the Milky Way~\cite{BovyRix}.
Its diffuse stellar disk exhibits a clear two-armed, grand design spiral pattern. This should be strongly suppressed by the dominant dark matter
halo: the disk-to-halo ratio is tiny, so there is insufficient disk self-gravity to drive the instabilities that feed spiral structure. And~yet, there it~is.

If MOND is the cause of spiral structure in LSB galaxies, then it is straightforward to predict~\cite{MdB98b}
how this would be interpreted in conventional terms.
Specifically, if~one were to apply the marginal stability condition~\cite{spiralmodes} to LSB galaxies, one would infer unnaturally 
large disk masses~\cite{MdB98b}. F568-1 (Figure \ref{Fuchs}) has a [3.6] luminosity of $6.3 \times 10^9\,\Lsun$ \cite{SPARC}, 
so population synthesis leads us to expect $M_* = 3.1 \times 10^9\,\Msun$.
In order to explain the observed spiral structure of F568-1, the~conventional analysis 
requires $M_* \approx 42 \times 10^9\,\Msun$ \cite{Fuchs_ASS}.
This is an order of magnitude more than expected for a normal stellar population, and~comparable to the 
much brighter Milky Way~\cite{BHGreview}.
Scaled to this mass, the~rotation curve of the disk accounts for essentially all the mass in this LSB galaxy (Figure \ref{Fuchs}b).
There is no room left for the dark matter halo, and~formally the disk exceeds the observed rotation at small radii.
Taken at face value, more mass is required to drive spiral structure than is allowed by the rotation curve.
This is the contradiction to conventional dynamics anticipated in the quote above~\cite{MdB98b}.

More generally, the~expectation is that the conventional analysis of spiral structure in LSB galaxies will indicate a stellar mass in excess
of that which is reasonable for stellar populations. This more general prediction is realized in the LSB galaxies for which a careful 
analysis has been performed~\cite{Fuchs_ASS,Fuchs_LSB,Saburova_LSB}. Quoting from these works: 
"these estimates seem to indicate that the disks of low surface brightness galaxies might be much more massive than currently thought. 
This puzzling result contradicts stellar population synthesis models" \cite{Fuchs_ASS}; 
"When I apply this method to the disks of low surface brightness galaxies, I find unexpectedly high mass-to light ratios" \cite{Fuchs_LSB};
and "For four low-surface-brightness galaxies, we find the disk masses corresponding to the marginal stability condition 
to be significantly higher than one may expect from their brightness" \cite{Saburova_LSB}.
These statements are exactly what was predicted to happen if disk stability is provided by MOND rather than a dark matter~halo.

It may be tempting to take these high masses literally rather than accept that they might be due to MOND.
Note, however, that all the correlations discussed above follow from stellar masses that are very much in accord with
our expectations for stellar populations. If~instead we adopt these higher masses, it will vastly increase the scatter in the BTFR
(Figure \ref{fig:BTFR}), building in a correlation of residuals with surface density where there are none with surface brightness,
and make nonsense of the correlations seen in Figures~\ref{VRSB} and \ref{ReSeAcc}. We cannot fix this without breaking~those.

\paragraph{\textit{Do the data corroborate the prediction of MOND?}} Yes. Application of modal analysis resulted in precisely
the predicted~effect.

\paragraph{\textit{Was the prediction made} a priori?} Yes. This was the obvious effect that could be anticipated for a conventional
analysis~\cite{MdB98b}. 

\paragraph{\textit{What does dark matter predict?}} Nominally, one expects LSB galaxies with stellar disk masses that are reasonable from the
perspective of stellar populations to be more stable than observed due to their tiny disk-to-halo ratios~\cite{MMdB}. 
There are any number of effects that could be invoked which might or might not circumvent this baseline expectation.
These do nothing to explain why the observed phenomenon follows from the predictions of~MOND.

\section{Discussion}
\label{sec:disc}

\begin{quote}
"In science, all new and startling facts must encounter in sequence the~responses
\begin{enumerate}
\item It is not true!
\item It is contrary to orthodoxy.
\item We knew it all along."
\end{enumerate}
---L.\ Agassiz (paraphrased)
\end{quote}

We have now completed each step in this progression.
The results of the earliest rotation curve data for LSB galaxies~\cite{dBMH96} were 
disputed\footnote{Systematic errors were repeatedly invoked. 
First it was beam smearing~\cite{beamsmearing}. This was a legitimate concern in a minority of cases;
it was addressed by improving the spatial resolution of the data with long slit observations~\cite{MRdB01,dBB02}. 
Then concerns were raised that these observations suffered from slit alignment errors~\cite{Swaters2003}. 
This was never a serious concern~\cite{MRdB01,dBBM03}, as~confirmed by subsequent improvements to the data~\cite{KdN06,KdN08}.
A variety of physical effects were then invoked; e.g.,~grossly non-circular motions~\cite{HNS07}, which could also be
excluded~\cite{KdN09,KdNK11}. This is what it looks like when the normal component makes excuses to disregard
the obvious implications of inconvenient data.} 
(it is not true!); attempts to pose the results in an empirical framework~\cite{M04} independent of MOND were met with 
antipathy\footnote{At the 2006 conference \textit{Galaxies in the Cosmic Web}, I showed~\cite{M04} that the systematic
dependence of the mass discrepancy on acceleration seen in Figure~\ref{MLSB}b was true empirically irrespective of
MOND. In~response, a~prominent galaxy formation theorist shouted, "We don't have to explain MOND!"} (it is against orthodoxy!),
and more recently, improved data~\cite{RAR} corroborating the results in~\cite{M04} 
are now frequently described\footnote{Recent papers describing the observed MONDian phenomenology as natural 
in \LCDM\ include~\cite{Ludlow2017,Navarro2017,RKKYPRX,Dutton_RAR,Coral_RAR}.   
If it were natural, it would have fallen out of \LCDM\ models long ago~\cite{MdB98a,MMW98}.}
as "natural" (we knew it all along).

Unfortunately, we did not know it all along. We have been surprised at every turn: these were startling facts, when new.
Only one theory succeeded in predicting these phenomena in advance: MOND. 
It has met the gold standard of scientific prediction repeatedly for a wide variety of phenomena, including many beyond the scope of
this review~\cite{LivRev,CJP,MM13a,MM13b,TucanaPrediction,Crater2}. I do not see how this can be a~fluke.


A common reaction at this juncture is "MOND may get X right, but~it gets\footnote{Most commonly, 
Y $=$ clusters of galaxies or large scale structure. These are discussed in~\cite{CJP} and references therein.} Y wrong. 
Therefore dark matter must be correct."
The second sentence does \textit{not} follow from the first, as~it presupposes that dark matter automatically explains everything that
MOND predicted in advance. This fails to address why MOND has predictive power that dark matter lacks:
just saying "dark matter does it" is not a satisfactory scientific explanation. We need to understand X irrespective of Y,
not use Y as an excuse to ignore~X.

The set X of properties discussed here is listed in Table~\ref{tab:mondpred}.
These include many successful \textit{a priori} predictions of MOND.
In contrast, many of these observations are problematic for the dark matter paradigm.
There is no good reason for properties (\ref{pr:masr}), (\ref{pr:accD}), (\ref{pr:fits}), and~(\ref{pr:freeman}) to arise in the context
of dark matter: they were not predicted, and~require fine-tuning to explain~\cite{M1999ASPC,MdB98a,MdB98b,M04,M05,myPRL,M11,M12,CJP}. 
Properties (\ref{pr:sbi}), (\ref{pr:renzo}), and~(\ref{pr:morph}) appear to be outright contradictions to the dark matter interpretation of galaxy dynamics. 
Whether they amount to a falsification depends on where we set that bar: what~would constitute a falsification of the dark matter? 
At the very least, it is disturbing that a completely different theory correctly predicted
a wide range of phenomena that the dark matter paradigm did~not.

\begin{table}[h]
\caption{MOND Predictions and Tests. \label{tab:mondpred}}
\centering
\begin{tabular}{lcc}
\toprule
\textbf{Prediction} & \textbf{Test Positive?}	 & \textbf{\textit{A Priori}?}\\
\midrule
\textbf{MASR (Tully--Fisher)} & & \\
Property \ref{pr:norm}. Normalization	& Yes	& No \\
Property \ref{pr:slope}.  Slope	& Yes	& No \\
Property \ref{pr:masr}.  Mass \& Asymptotic Speed	& Yes	& Yes \\
Property \ref{pr:sbi}.  Surface Brightness Independence	& Yes	& Yes \\
\textbf{Rotation Curves} & & \\
Property \ref{pr:flat}.  Flat Rotation Curves 	& Yes 	& No \\
Property \ref{pr:accD}.  Acceleration Discrepancy		& Yes	& Yes \\
Property \ref{pr:shape}.  Rotation Curve Shapes		& Yes	& Yes \\
Property \ref{pr:sdens}.  Surface Brightness \& Density	& Yes	& Yes \\
Property \ref{pr:fits}.  Detailed Fits 	& Yes	& No \\
Property \ref{pr:pops}.  Stellar Population \ML\	& Yes	&---\\
Property \ref{pr:renzo}.  Feature Correspondence & Yes &---\\
\textbf{Disk Stability} & & \\
Property \ref{pr:freeman}.  Freeman Limit	& Yes	& No \\
Property \ref{pr:vvd}.  Vertical Velocity Dispersions 	& {?} 	& No \\
Property \ref{pr:morph}.  LSB Galaxy Morphology	& Yes	& Yes \\
\bottomrule
\end{tabular}
\end{table}

In most cases, the~MOND-predicted properties in Table~\ref{tab:mondpred} are obvious in the data with no \textit{fitting} whatsoever.
For example, of~the four properties of the MASR, only its normalization must be fit.
Once the value of \azero\ is specified, the~slope is fixed, and~is consistent with subsequently obtained data.
That the MASR would be independent of surface brightness was also a genuine, and~conventionally unexpected, \textit{a priori}
prediction. That the relation was fundamentally one between baryonic mass and \Vf\ was first anticipated by~MOND. 

In a similar manner, many of the predicted properties of rotation curves follow directly without recourse to fitting. This includes 
the amplitude of the acceleration discrepancy (Figure \ref{MLSB}), the~shapes of rotation curves (Figure \ref{VRSB}), and~the dependence
of acceleration on surface brightness (Figure \ref{ReSeAcc}). The~predictions of MOND can be seen directly in the~data.

In order to make detailed rotation curve fits, we must treat the stellar mass-to-light ratio as a fit parameter. 
This one degree of freedom is unavoidable in any theory. MOND fits work well with a single, 
universal value of \azero\ \cite{S1996,SV1998,LiRAR}. The~value of \azero\ is not allowed to vary from galaxy to galaxy, and~there is no
indication\footnote{The literature contains contradictory statements on this point~\cite{Rodrigues18}, 
but these usually stem from holding MOND to a higher standard than dark matter. 
Galaxies that have bad fits in MOND also have bad dark matter fits (in terms of $\chi_{\nu}^2$ \cite{SPARChalos}). 
This is a sign that the uncertainties have been underestimated, not that \azero\ must vary or that all conceivable models are wrong.} 
in the data of a need to do so~\cite{LiRAR}.
An independent test of the best-fit values of \ML\ is provided by stellar population synthesis models.
The agreement with these the two could hardly be better (Figure \ref{MLSBpop}).

When the cold dark matter paradigm became widely accepted, the~only properties in Table~\ref{tab:mondpred} that had seriously
informed its development was that rotation curves are flat and the Oort discrepancy exists. 
These were taken to mean that there had to be dark matter, and~little more.
The Tully--Fisher relation was known at the time, but~was widely viewed as a method to determine distances, not inform theory.
The remaining elements of Table~\ref{tab:mondpred} were essentially unknown, or, in~the case of Freeman's Law, widely misinterpreted.
It is not obvious that we would develop the same paradigm had we known then what we know now.

\section{Conclusions}
\label{sec:conc}

Many predictions of MOND have been corroborated over the years. {It has repeatedly met the gold standard 
of the scientific method in which predictions are made in advance of their observation.
The~dark matter paradigm does not share a comparable record of predictive success in galaxy dynamics.}

There are three broad categories of interpretation admitted by the data discussed~here.
\begin{enumerate}
\item The data corroborate the predictions of MOND because there is something to it.
\item The physics of galaxy formation somehow mimic MOND, at~least for rotating galaxies.
\item There is something new and different going on that we have yet to imagine.
\end{enumerate}

These are essentially identical to the possibilities discussed over 20 years ago~\cite{MdB98b}, 
with the addition of (3), which is sufficiently vague to always be a logical possibility. 
There has been some progress in this direction, with~hypotheses for dark fluids~\cite{ZB2010}, or~bipolar~\cite{bipolarDM} or 
superfluid~\cite{superfluidDM} dark matter with built-in MOND-like behavior while retaining the putative successes of CDM on large scales. 
There remains a great deal to be explored in this~direction. 

Nevertheless, the~obvious interpretation of the data discussed here is (1): MOND gets all these predictions correct, in~advance of their
observation, because~there is something to it. This motivates the search for a satisfactory theory that encompasses both general relativity 
and MOND. Some progress has been made along these lines~\cite{TeVeS,bimetric,LivRev,milgromrev2014,Clifton12,SkordisZlosnik19,HSN2019},
but overall, shockingly little effort has been made to investigate in this~possibility.

In contrast, an~\textit{enormous} amount of effort has been invested in (2),
a thorough discussion of \textbf{which} is well beyond the scope of this review.
{However, the~basic problem is simple: MOND has made many successful, \textit{a priori} predictions that dark matter did not.
We are obliged to adjust our dark matter models to accommodate the successful predictions of a contrary theory.}

There remains a considerable amount that we do not understand about the universe,
including whether the invisible particles hypothesized to dominate its mass budget actually~exist.

\begin{quote}
"The normal component (i.e., the~accepted paradigm and its adherents) is large and well entrenched. Hence, a~change of the normal component is very noticeable. So is the resistance of the normal component to change. This resistance becomes especially strong and noticeable in periods where a change seems to be~imminent." \\
---P.\ Feyerabend~\cite{Feyerabend}
\end{quote}


\vspace{6pt} 




\funding{This research has been supported in part by NSF grant PHY-1911909 and NASA grant~80NSSC19K0570.}

\acknowledgments{The author is grateful to many colleagues for critical conversations over the years; in~particular,
Greg Bothun, Jim Schombert, Doug Richstone, Joel Bregman, Jim Peebles, Houjun Mo, Simon White, Donald Lynden-Bell, 
Chris Mihos, Thijs van der Hulst, Renzo Sancisi, Erwin de Blok, Bob Sanders, Vera~Rubin, Moti Milgrom, Jerry Sellwood, 
David Merritt, Arthur Kosowsky, James Binney, Frank van den Bosch, Roelof de Jong, Eric Bell, Rachel Kuzio de Naray, 
Benoit Famaey, Pavel Kroupa, Marcel Pawlowski, Federico Lelli, Hongsheng Zhao, and~Xavier~Hernandez.}

\conflictsofinterest{The author declares no conflict of~interest.} 

\abbreviations{The following abbreviations are used in this manuscript:\\

\noindent 
\begin{tabular}{@{}ll}
BTFR & Baryonic Tully--Fisher relation \\
CDM & Cold dark matter \\
HSB & High surface brightness\\
IMF & Initial mass function \\
\LCDM\ & Lambda cold dark matter \\
LSB & Low surface brightness\\
MASR & Mass--asymptotic speed relationn \\
MOND & Modified Newtonian dynamics
\end{tabular}}

\reftitle{References}

\end{document}